\documentclass[8.5pt,twoside,twocolumn,amsmath,amssymb]{article}
\usepackage[super,sort&compress,comma]{natbib}
\usepackage{mhchem}
\usepackage{times,mathptmx}
\usepackage{sectsty}
\usepackage{balance}

\usepackage{graphicx} 
\usepackage{lastpage}
\usepackage[format=plain,singlelinecheck=false,font=small,labelfont=bf,labelsep=space]{caption}
\usepackage{fancyhdr}
\usepackage{dcolumn}
\usepackage{bm}
\usepackage{amsmath}
\usepackage{bbding}
\usepackage{verbatim}

\newcommand{\etalia}{{\it et al.~}}
\newcommand{\la}{\left\langle}
\newcommand{\ra}{\right\rangle}
\newcommand{\bla}{\Big\langle}
\newcommand{\bra}{\Big\rangle}
\newcommand{\kt}{\tilde\kappa}
\newcommand{\rt}{\tilde r}

\newcommand{\PRL}{Phys.~Rev.~Lett.}

\begin{document}

\thispagestyle{plain}
\fancypagestyle{plain}{
\fancyhead[L]{\includegraphics[height=8pt]{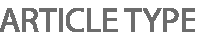}}
\fancyhead[C]{\hspace{-1cm}\includegraphics[height=20pt]{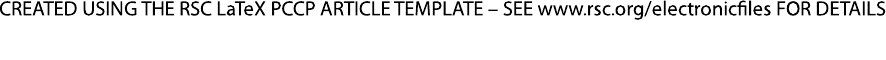}}
\fancyhead[R]{\includegraphics[height=10pt]{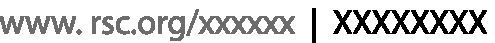}\vspace{-0.2cm}}
\renewcommand{\headrulewidth}{1pt}}
\renewcommand{\thefootnote}{\fnsymbol{footnote}}
\renewcommand\footnoterule{\vspace*{1pt}%
\hrule width 3.4in height 0.4pt \vspace*{5pt}}
\setcounter{secnumdepth}{5}

\makeatletter
\def\subsubsection{\@startsection{subsubsection}{3}{10pt}{-1.25ex plus -1ex minus -.1ex}{0ex plus 0ex}{\normalsize\bf}}
\def\paragraph{\@startsection{paragraph}{4}{10pt}{-1.25ex plus -1ex minus -.1ex}{0ex plus 0ex}{\normalsize\textit}}
\renewcommand\@biblabel[1]{#1}
\renewcommand\@makefntext[1]%
{\noindent\makebox[0pt][r]{\@thefnmark\,}#1}
\makeatother
\renewcommand{\figurename}{\small{Fig.}~}
\sectionfont{\large}
\subsectionfont{\normalsize}

\fancyfoot{}
\fancyfoot[LO,RE]{\vspace{-7pt}\includegraphics[height=9pt]{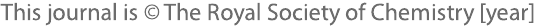}}
\fancyfoot[CO]{\vspace{-7.2pt}\hspace{12.2cm}\includegraphics{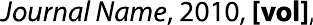}}
\fancyfoot[CE]{\vspace{-7.5pt}\hspace{-13.5cm}\includegraphics{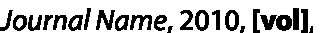}}
\fancyfoot[RO]{\footnotesize{\sffamily{1--\pageref{LastPage} ~\textbar  \hspace{2pt}\thepage}}}
\fancyfoot[LE]{\footnotesize{\sffamily{\thepage~\textbar\hspace{3.45cm} 1--\pageref{LastPage}}}}
\fancyhead{}
\renewcommand{\headrulewidth}{1pt}
\renewcommand{\footrulewidth}{1pt}
\setlength{\arrayrulewidth}{1pt}
\setlength{\columnsep}{6.5mm}
\setlength\bibsep{1pt}

\twocolumn[
  \begin{@twocolumnfalse}
\noindent\LARGE{\textbf{Concentration-Dependent Swelling and Structure of Ionic Microgels: \\
Simulation and Theory of a Coarse-Grained Model}}
\vspace{0.6cm}

\noindent\large{\textbf{Tyler J.~Weyer and Alan R.~Denton\textit{$^{\ast}$}
}}\vspace{0.5cm}

\noindent\textit{\small{\textbf{Received 10th May 2018, Accepted Xth XXXXXXXXX 20XX\newline
First published on the web Xth XXXXXXXXXX 200X}}}

\noindent \textbf{\small{DOI: 10.1039/b000000x}}
\vspace{0.6cm}

\noindent \normalsize{
We study swelling and structural properties of ionic microgel suspensions within a comprehensive 
coarse-grained model that combines the polymeric and colloidal natures of microgels as permeable, 
compressible, charged spheres governed by effective interparticle interactions.  The model synthesizes 
the Flory-Rehner theory of cross-linked polymer gels, the Hertz continuum theory of effective elastic 
interactions, and a theory of density-dependent effective electrostatic interactions.  Implementing 
the model using Monte Carlo simulation and thermodynamic perturbation theory, we compute equilibrium 
particle size distributions, swelling ratios, volume fractions, net valences, radial distribution functions,
and static structure factors as functions of concentration.  Trial Monte Carlo moves comprising particle 
displacements {\it and} size variations are accepted or rejected based on the total change in 
elastic and electrostatic energies.  The theory combines first-order thermodynamic perturbation and 
variational free energy approximations.  For illustrative system parameters, theory and simulation 
agree closely at concentrations ranging from dilute to beyond particle overlap.  With increasing 
concentration, as microgels deswell, we predict a decrease in the net valence and an unusual saturation 
of pair correlations.  Comparison with experimental data for deionized, aqueous suspensions of 
PNIPAM particles demonstrates the capacity of the coarse-grained model to predict and interpret 
measured swelling behavior.
}
\vspace{0.6cm}
 \end{@twocolumnfalse}
  ]

\footnotetext{\textit{Department of Physics, North Dakota State University, 
Fargo, ND 58108-6050, USA.  E-mail: alan.denton@ndsu.edu}}

\section{Introduction}

Microgels are soft colloidal particles, composed of cross-linked polymer gels,
possessing internal degrees of freedom that allow them to swell to many times their 
dry size when dispersed in a solvent.\cite{baker1949,pelton1986,pelton2000,saunders2009}
Porosity and compressibility enable microgels to adjust their size in response to 
changes in temperature, $p$H, and concentrations of different species.
Responsiveness to environmental conditions, coupled with ability to absorb and transport 
cargo, e.g., fluorescent dye or drug molecules, facilitates applications of microgels to 
biosensing and drug delivery.  
\cite{HydrogelBook2012,MicrogelBook2011,lyon-nieves-AnnuRevPhysChem2012,yunker-yodh-review2014}
In a polar solvent, microgels may acquire charge (ionize) via dissociation 
of counterions into solution.  Salt ions, whether naturally present or added, contribute 
to the total population of free microions (counterions and coions), which screen the bare 
Coulomb interactions between ionic microgels.

The elastic properties of microgels have been explored in numerous experimental and modeling studies.
\cite{
cloitre-leibler1999,
cloitre-leibler2003,
tan2004,
nieves-macromol2000,
nieves-jcp2003,
nieves-macromol2009,
hellweg2010,
weitz-sm2012,
weitz-jcp2012,
ciamarra2013,
nieves-sm2011,
schurtenberger-ZPC2012,
nieves-sm2012,
nieves-bulk-shear-pre2011,
nieves-bulk-pre2011,
dufresne2009,
nieves-prl2015}
Experimental measurements of microgel swelling have deployed an array of techniques,
including static and dynamic light scattering, optical microscopy, small-angle neutron 
scattering, and osmometry.  
\cite{mohanty-richtering2008,richtering2008,lyon2007,weitz-pre2012,
schurtenberger-SM2012,holmqvist-shurtenberger2012,holmqvist-shurtenberger2012-erratum,
schurtenberger2013,nieves-pre2013,schurtenberger2014,braibanti-perez2016}
Suspensions of soft microgels display thermodynamic, structural, and dynamical properties 
that differ significantly from those of suspensions of hard colloids.
\cite{
weitz-prl1995,
groehn2000,
levin2002,
nieves-jcp2005,
winkler-gompper2012,
winkler-gompper2014,
winkler2017,
zaccarelli2017,
li-chen2014,
egorov-likos2013,
colla-likos2014,
colla-likos2015,
colla-likos2018,
stellbrink-likos-nanoscale2015,
stellbrink-likos-prl2015}
Differences in bulk properties are tied to single-particle compressibility and swelling,
which are governed by polymer gel elasticity and entropy, polymer-solvent interactions, and 
-- in the case of ionic microgels -- electrostatic interactions.  

Despite many studies, the complex interplay between elastic and electrostatic influences on the 
swelling behavior and bulk properties of ionic microgels is still not widely appreciated and is 
only partially understood.
In previous work on thermodynamic and structural properties of microgel suspensions, we modeled 
ionic microgels as charged, elastic, but incompressible spheres\cite{hedrick-chung-denton2015} 
and nonionic microgels as uncharged, elastic, and compressible spheres.\cite{urich-denton2016}  
Here we combine these approaches to model ionic microgels as charged, elastic, and compressible spheres.
The purpose of this paper is to analyze the combined influences of particle compressibility 
and elastic and electrostatic interparticle interactions on equilibrium thermal and structural 
properties of ionic microgel suspensions.  

The outline of the paper is as follows.  In Sec.~\ref{models}, we derive a coarse-grained
one-component model of ionic microgel suspensions, in which the microgels are represented as 
compressible, charged spheres and the solvent and microions appear implicitly through 
effective interparticle interactions.  The model is a synthesis of a single-particle 
polymer free energy, approximated via the Flory-Rehner theory of cross-linked polymer 
networks,\cite{flory-rehner1943-I,flory-rehner1943-II,flory1953} an electrostatic self energy,
and interparticle interactions, approximated by combining a Hertz elastic pair 
potential\cite{landau-lifshitz1986} with effective electrostatic interactions, derived 
from Poisson-Boltzmann theory.\cite{denton2003}  In Sec.~\ref{methods}, we describe
two computational methods -- Monte Carlo simulation and thermodynamic perturbation theory --
with which we implemented the coarse-grained model to predict properties of ionic microgel
suspensions.  Section~\ref{results} presents results for the equilibrium swelling ratio,
volume fraction, net valence, radial distribution function, and static structure factor 
as functions of concentration.  Section~\ref{conclusions} summarizes and concludes.

\section{Models}\label{models}

\subsection{Ionic Microgels}\label{microscopic-model}

\begin{figure}
\includegraphics[width=\columnwidth]{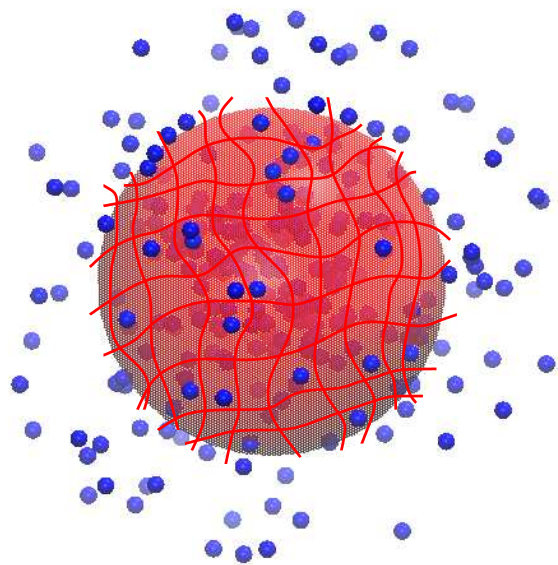}
\vspace*{-0.2cm}
\caption{
Schematic drawing of an ionic microgel (large red sphere), with cross-linked polymer chains shown
to suggest internal structure, surrounded by oppositely charged counterions (small blue spheres).
In coarse-grained model, microions and polymer chains are implicit.
}\label{fig1}
\end{figure}

An ionic microgel is a soft colloidal particle, consisting of a cross-linked polymer network 
swollen by a polar solvent, from which ions (counterions) have dissociated (see Fig.~\ref{fig1}).  
Depending on the chemical synthesis, the counterions may originate from the polymer chains or 
from the initiator in the polymerization.\cite{lyon-langmuir2011}
While the distributions of monomers, cross-linkers, and fixed charges depend on the synthesis method,
we assume for simplicity uniform average distributions.
This reference model can be generalized to heterogeneous microgels with core-shell or hollow
structures.\cite{stieger2004,nieves-sm2011,weitz-jcp2012,schurtenberger-ZPC2012,ciamarra2013,
moncho-jorda-anta2013,boon-schurtenberger2017,quesada-perez2013,quesada-perez-moncho-jorda-anta2015,
potemkin2015}  Assuming random close packing of monomers in the unswollen (dry) state, the dry 
particle radius $a_0$ is determined by the number $N_{\rm mon}$ and radius $a_{\rm mon}$ 
of monomers making up a particle via $a_0\simeq(N_{\rm mon}/0.63)^{1/3}a_{\rm mon}$.
The swollen particle radius $a$ is determined by the fraction of monomers that are 
cross-linked, the number of fixed charges (valence), and the solution conditions,
including temperature, solvent quality, and concentrations of salt and microgels. 

We consider $N_m$ spherical microgels of valence $Z$ dispersed in a solvent of volume $V$ 
with $N_{\pm}$ counterions/coions.  Assuming negatively charged microgels (charge $-Ze$),
global electroneutrality requires $ZN_m=N_+-N_-$, which equals the number of counterions that dissociate 
from the polymer chains.  In a closed system, the number of salt ion pairs $N_s=N_-$ is fixed.
In the case of Donnan equilibrium between the suspension and an electrolyte reservoir across
a semipermeable membrane, the salt concentration of the suspension is determined by the 
reservoir concentration.
Given the average microgel number density $n_m=N_m/V$, the dry volume fraction, $\phi_0=(4\pi/3)n_m a_0^3$,
is defined as the fraction of the total volume occupied by the particles in their dry state.  
For swollen particles of radius $a$ (swelling ratio $\alpha=a/a_0$), the actual volume fraction,
$\phi=(4\pi/3)n_m a^3=\phi_0\alpha^3$, can substantially exceed $\phi_0$ and may even exceed the 
close-packing limit for hard spheres if particles become faceted or otherwise deform in shape.
\cite{riest2015,cloitre-bonnecaze2010,cloitre-bonnacaze2011} 

\subsection{Coarse-Grained Model}\label{coarse-grained-model}

Rather than attempt to explicitly model all monomers and ions in a microgel suspension, we 
develop a more computationally practical approach that averages over solvent, polymer, and 
microion degrees of freedom to reduce a suspension of ionic microgels to a system of
elastic, charged, spherical particles governed by an effective Hamiltonian that comprises 
both a single-particle free energy and effective interparticle interactions.  The resulting 
coarse-grained model, incorporating both the polymeric and the colloidal natures of microgels,
can be derived from a molecular model by first tracing out the solvent degrees of freedom,
leading to the primitive model, with the solvent replaced by a dielectric continuum, and then 
tracing out the remaining microscopic (polymer and microion) degrees of freedom, leading to 
a one-component model with microgels replaced by pseudo-microgels.  

In the canonical ensemble, with fixed numbers of particles in a volume $V$ at temperature $T$, 
the partition function of the system in the primitive model may be expressed as
\begin{equation}
{\cal Z}=\bla\bla\bla e^{-\beta(K+H_m+H_{mm}+H_{m\mu}+H_{\mu\mu})} \bra_p\bra_{\mu}\bra_m,
\label{Z}
\end{equation}
where $\beta\equiv 1/(k_BT)$, $K$ is the total kinetic energy of the system,
and angular brackets denote traces over polymer ($p$), microion ($\mu$), and center-of-mass 
microgel ($m$) coordinates.  The polymer coordinates are internal degrees of freedom of the 
microgels associated with motion of segments making up the cross-linked polymer chains.
In the Boltzmann factor, $H_m$ is the single-microgel Hamiltonian, comprising both polymeric and 
electrostatic self energies, $H_{mm}$ incorporates polymeric and electrostatic energies of interaction 
between microgels, and $H_{m\mu}$ and $H_{\mu\mu}$ account, respectively, for 
microgel-microion and microion-microion interactions.  

Since classical traces commute, the trace over polymer coordinates can be independently performed,
with the result
\begin{equation}
{\cal Z}=e^{-\beta(U_e+F_p)}\bla\bla e^{-\beta(K+H_{mm}+H_{m\mu}+H_{\mu\mu})} \bra_{\mu}\bra_m,
\label{Z1}
\end{equation}
where, if we assume spherical microgels of swollen radii $a_i$ ($i=1,\ldots,N_m$),
\begin{equation}
U_e=\sum_{i=1}^{N_m}\, u_e(a_i)
\label{Ue}
\end{equation}
is the electrostatic self energy of the fixed charges and 
\begin{equation}
F_p=\sum_{i=1}^{N_m}\, f_p(a_i)
\label{Fp}
\end{equation}
is the free energy associated with polymeric degrees of freedom within the microgels.
For uniformly distributed fixed charges, the single-microgel electrostatic self energy is
\begin{equation}
u_e(a)=\frac{3}{5}\frac{Z^2e^2}{\epsilon a},
\label{ue}
\end{equation}
where $\epsilon$ is the dielectric constant of the implicit solvent.  To approximate the 
single-microgel polymer free energy, we adopt the Flory-Rehner theory of polymer networks.
\cite{flory-rehner1943-I,flory-rehner1943-II,flory1953}  In the case of uniformly distributed 
cross-linkers that divide the network into $N_{\rm ch}$ chains, the Flory-Rehner theory predicts 
\begin{eqnarray}
\beta f_p(\alpha)&=&N_{\rm mon}\left[(\alpha^3-1)\ln\left(1-\alpha^{-3}\right) 
+\chi\left(1-\alpha^{-3}\right)\right] 
\nonumber\\[1ex]
&+&\frac{3}{2}N_{\rm ch}\left(\alpha^2-\ln\alpha-1\right),
\label{fp}
\end{eqnarray}
where $\chi$ is the polymer-solvent interaction (solvency) parameter.  
In Eq.~(\ref{fp}), the first term in square brackets combines the entropy of mixing of 
microgel monomers and solvent molecules with a mean-field approximation 
for the polymer-solvent interaction, which neglects interparticle correlations.  
The last term in Eq.~(\ref{fp}) accounts for the elastic free energy of stretching the 
microgel network by assuming isotropic deformation, ignoring changes in internal energy
associated with the structure of the surrounding solvent, and modeling polymers as Gaussian coils.
The Gaussian model is reasonable for chain end-to-end displacements much shorter than 
the polymer contour length,\cite{potemkin2015} which implies swelling ratios
$\alpha\ll \sqrt{N_{\rm mon}/N_{\rm ch}}$.
Numerous studies 
\cite{weitz-jcp2012,
schurtenberger-ZPC2012,
ciamarra2013,
colla-likos2014,
colla-likos2015,
nieves-sm2012,
weitz-sm2012,
nieves-bulk-shear-pre2011,
nieves-sm2011,
nieves-bulk-pre2011,
nieves-macromol2009,
nieves-jcp2003,
nieves-macromol2000,
pelton2000,
hellweg2010,
saunders2009,
moncho-jorda-dzubiella2016}
have established that the Flory-Rehner theory, although originally developed for macroscopic gels,
yields a reasonable description of the elastic properties of loosely cross-linked microgels, 
despite overestimating the solvency parameter.\cite{richtering2017}
Nevertheless, more realistic and accurate theories of polymer network swelling could be 
incorporated into the model.\cite{blundell2009,dzubiella2017}

Returning to the trace over microgel and microion coordinates in Eq.~(\ref{Z1}), the term $H_{mm}$ 
represents the total internal energy associated with effective elastic and bare (Coulomb) 
electrostatic interactions between microgels.  If we assume pairwise additive elastic forces, 
a practical model of effective elastic interactions is provided by 
the Hertz potential,\cite{landau-lifshitz1986}
\begin{equation}
v_H(r)=\left\{ \begin{array} 
{l@{\quad\quad}l}
B_{ij}\left(1-\frac{\displaystyle r}{\displaystyle a_i+a_j}\right)^{5/2},
& r\leq a_i+a_j \\[1ex]
0~, 
& r>a_i+a_j, \end{array} \right.
\label{Hertz}
\end{equation}
whose amplitude depends on the elastic properties of the gel through 
Young's modulus $Y_i$ and Poisson's ratio $\nu_i$\cite{landau-lifshitz1986}:
\begin{equation}
B_{ij}=\frac{8}{15}\left(\frac{1-\nu_i^2}{Y_i}+\frac{1-\nu_j^2}{Y_j}\right)^{-1}
(a_i+a_j)^2\sqrt{a_i a_j}.
\label{B1}
\end{equation}
In the case of equal radii ($a$) and equal elastic constants ($Y$, $\nu$),
the Hertz amplitude simplifies to
\begin{equation}
B=\frac{16Y a^3}{15(1-\nu^2)}.
\label{B2}
\end{equation}
For polymer gels in good solvents, scaling theory\cite{deGennes1979} predicts that Young's modulus 
scales linearly with temperature and cross-linker number density: $Y\sim TN_{\rm ch}/a^3$.  
Thus, the reduced Hertz amplitude, $B^*\equiv\beta B$, is proportional to $N_{\rm ch}$ and is essentially 
independent of temperature and particle volume, neglecting any dependence of $\nu$ on $\alpha$. 
The total internal energy associated with pair interactions is then approximated by
\begin{equation}
U_{mm}=\sum_{i<j=1}^{N_m}\, [v_H(r_{ij})+v_C(r_{ij})],
\label{Umm}
\end{equation}
where $r_{ij}$ is the center-to-center separation of particles $i$ and $j$
and $v_C(r)$ is the bare Coulomb interaction between microgels.

The coarse-graining procedure is completed by tracing out the microion degrees of freedom.
This step reduces the partition function of the multi-component mixture to that of a 
one-component model (OCM) of pseudo-microgels,
\begin{equation}
{\cal Z}_{\rm OCM}=\bla e^{-\beta H_{\rm eff}}\bra_m,
\label{Z2}
\end{equation}
governed by an effective Hamiltonian,
\begin{equation}
H_{\rm eff}=K_m+U_e+F_p+U_{mm}-k_BT\ln\bla e^{-\beta(K_{\mu}+H_{m\mu}+H_{\mu\mu})}\bra_{\mu},
\label{Heff1}
\end{equation}
which involves effective electrostatic interactions between the pseudo-microgels.
Here $K_m$ and $K_{\mu}$ are the kinetic energies of the microgels and microions, respectively. 
If the microion densities respond linearly to the electrostatic potential of the microgels, 
the effective electrostatic interactions are limited to one- and two-body contributions.
\cite{denton2003,denton2000,denton-cecam2014}
Under this approximation, the effective Hamiltonian takes the form
\begin{equation}
H_{\rm eff}=K_m+U_e+F_p+E_V(n_m)+U_{\rm eff}(n_m),
\label{Heff2}
\end{equation}
where $E_V(n_m)$ is a one-body volume energy and 
\begin{equation}
U_{\rm eff}(n_m)=\sum_{i<j=1}^{N_m}\, \left[v_H(r_{ij})+v_{\rm eff}(r_{ij}; n_m)\right]
\label{Ueff}
\end{equation}
is the microgel-microgel interaction energy, which combines the Hertz elastic pair potential 
with an effective electrostatic pair potential $v_{\rm eff}(r; n_m)$.  Note that $E_V(n_m)$ and
$v_{\rm eff}(r; n_m)$ both depend on the average microgel density
and consistently incorporate screening of the fixed network charge by mobile microions.
Equations (\ref{Heff2}) and (\ref{Ueff}) constitute a formal expression of the coarse-grained 
model of ionic microgels.  Although we adopt here the Flory-Rehner and Hertz models for 
$f_p$ and $v_H(r)$, alternative models of polymer gels could be substituted.

Still required for applications is specification of the effective electrostatic interactions.
In previous work,\cite{denton2003} one of us combined the linear-response approximation 
with a random-phase approximation for the response functions of the microion plasma 
to derive practical expressions for the effective interactions.
The form of the effective electrostatic pair potential depends on whether or not
the microgels are overlapping:  
\begin{equation}
v_{\rm eff}(r)= 
\left\{ \begin{array} {l@{\quad\quad}l}
v_{\scriptscriptstyle Y}(r), & r>a_i+a_j
\\[2ex]
v_{\rm ov}(r), & r\leq a_i+a_j.
\end{array} \right. 
\label{veff}
\end{equation}
Nonoverlapping microgels ($r>a_i+a_j$) interact via an effective Yukawa (screened-Coulomb) pair potential,
\begin{equation}
\beta v_{\scriptscriptstyle Y}(r)=\lambda_B Z_{\rm net}(a_i)Z_{\rm net}(a_j)
\frac{\displaystyle e^{\kappa(a_i+a_j)}}{\displaystyle (1+\kappa a_i)(1+\kappa a_j)}
\frac{\displaystyle e^{-\kappa r}}{\displaystyle r}, 
\label{vY}
\end{equation}
where
\begin{equation}
\kappa=\sqrt{4\pi\lambda_B(n_++n_-)}=\sqrt{4\pi\lambda_B(Zn_m+2n_s)}
\label{kappa}
\end{equation}
is the Debye screening constant, which depends on the average microgel density and the 
salt ion pair density $n_s=N_s/V$, and
\begin{equation}
Z_{\rm net}(a)=(1+\kappa a)e^{-\kappa a}
\frac{3Z}{\kappa^2 a^2}\left(\cosh(\kappa a)-\frac{\sinh(\kappa a)}{\kappa a}\right)
\label{znet}
\end{equation}
is the linear-response theory prediction\cite{denton2003} for the net valence of a microgel, 
defined as the bare valence $Z$ times the fraction of counterions exterior to the microgel.
In passing, we note that, since $Z_{\rm net}$ depends on the product $\kappa a$,
and since $\kappa$ increases, while $a$ decreases, with increasing microgel concentration,
$Z_{\rm net}$ depends nontrivially on concentration and swelling ratio.

For overlapping microgels, the effective electrostatic pair potential can be decomposed as
\begin{equation}
v_{\rm ov}(r)=v_{mm}(r)+v_{\rm ind}(r), \quad r\leq a_i+a_j,
\label{vr<2a-gel}
\end{equation}
where $v_{mm}(r)$ is the bare (Coulomb) pair potential and $v_{\rm ind}(r)$ is the microion-induced 
potential.  In the simplest case of uniformly charged microgels of equal size,\cite{denton2003}
\begin{equation}
\beta v_{mm}(r)=Z^2\frac{\lambda_B}{a}\left(\frac{6}{5}-\frac{1}{2}\rt^2
+\frac{3}{16}\rt^3-\frac{1}{160}\rt^5\right)
\label{vmmr<2a-gel}
\end{equation}
and
\begin{eqnarray}
&&\beta v_{\rm ind}(r)=
-\left(\frac{3Z}{\kt^2}\right)^2\frac{\lambda_B}{2r}\left[
\left(1+\frac{1}{\kt}\right)^2e^{-2\kt}\sinh(\kappa r)\right.
\nonumber\\[1ex]
&&+\left(1-\frac{1}{\kt^2}\right)
\left(1-e^{-\kappa r}+\frac{1}{2}\kappa^2r^2+\frac{1}{24}\kappa^4r^4\right)
\nonumber\\[1ex]
&&-\left.\frac{2}{3}\kt^2\left(1-\frac{2}{5}\kt^2\right)\rt
-\frac{1}{9}\kt^4\rt^3-\frac{1}{720}\kt^4\rt^6 \right],
\label{vindr<2a-gel}
\end{eqnarray}
with $\kt\equiv \kappa a$ and $\rt\equiv r/a$.
We omit the generalizations of Eqs.~(\ref{vmmr<2a-gel}) and (\ref{vindr<2a-gel}) to microgels 
of different sizes, as they are not needed in the applications considered below in
Sec.~\ref{results}.

Within the same approximations, the volume energy takes the explicit form
\begin{eqnarray}
&&
\hspace*{-1cm}
\beta E_V=\beta F_{\rm plasma}-3\lambda_B Z^2\sum_{i=1}^{N_m}\, \frac{1}{a_i}
\left\{\frac{1}{5}-\frac{1}{2(\kappa a_i)^2}\right.
\nonumber\\[1ex]
&&
\hspace*{-1cm}
+\frac{3}{4(\kappa a_i)^3}
\left.
\left[1-\frac{(1+\kappa a_i)^2}{(\kappa a_i)^4}
e^{-2\kappa a_i}\right]\right\}
-\frac{ZN_m}{2}\frac{n_+-n_-}{n_++n_-},
\label{EV}
\end{eqnarray}
where $n_{\pm}=N_{\pm}/V$ are the average microion densities and
\begin{equation}
\beta F_{\rm plasma}~=~N_+[\ln(n_+\Lambda^3)-1]+N_-[\ln(n_-\Lambda^3)-1]
\label{fplasma}
\end{equation}
is the ideal-gas free energy of a plasma of microions in a uniform compensating background,
$\Lambda$ being the thermal de Broglie wavelength.  
Equations~(\ref{vY}) and (\ref{EV}) are straightforward generalizations of the previously derived 
effective electrostatic interactions\cite{denton2003} to polydisperse suspensions.  
\cite{colla-likos2015}

The coarse-grained one-component model developed above synthesizes previously studied models of 
incompressible, ionic microgels\cite{denton2003,hedrick-chung-denton2015} and compressible, nonionic 
microgels.\cite{urich-denton2016}  Within this comprehensive model, particle swelling is determined by 
(1) elastic free energy of the polymer network internal to the microgels, approximated by the 
Flory-Rehner free energy [Eq.~(\ref{fp})]; 
(2) elastic interparticle interactions, approximated by the Hertz potential [Eq.~(\ref{Hertz})]; 
(3) electrostatic self energy of the microgels [Eq.~(\ref{ue})]; and 
(4) effective electrostatic interactions between microgels, approximated by a
linear-response theory [Eqs.~(\ref{veff})-(\ref{fplasma})].  
In the next section, we describe computational methods for implementing the model.

\section{Methods}\label{methods}

\subsection{Monte Carlo Simulation}\label{MC}
\begin{figure}
\includegraphics[width=0.9\columnwidth]{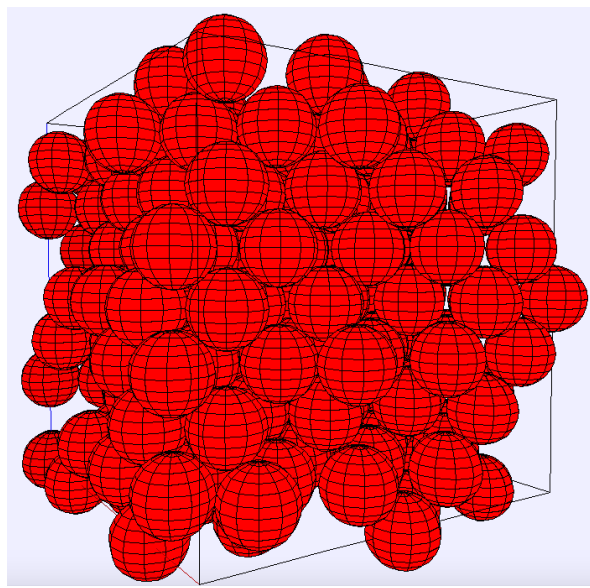}
\vspace*{-0.2cm}
\caption{
Typical snapshot from a simulation of a suspension of compressible, ionic, spherical microgels 
of fluctuating size in a cubic box with periodic boundary conditions in the coarse-grained model.
}\label{fig2}
\end{figure}

To predict equilibrium swelling behavior and thermal and structural properties of bulk suspensions 
of ionic microgels, we developed a Monte Carlo (MC) simulation method suited to the coarse-grained 
one-component model described in Sec.~\ref{models}.  We performed constant-$NVT$ (canonical ensemble)
simulations of pseudo-microgels confined to a cubic cell with periodic boundary conditions at 
fixed system parameters: $\lambda_B$, $Z$, $a_0$, $N_{\rm mon}$, $N_{\rm ch}$, $\chi$, $B^*$, 
$\phi_0$, and $n_s$.  Figure~\ref{fig2} shows a typical snapshot of the system.
In a variation of the conventional Metropolis algorithm,\cite{frenkel-smit2001,binder-heermann2010} 
our method involves trial moves that combine both displacements {\it and} changes in size of the 
particles.  \cite{urich-denton2016}  A trial move that simultaneously displaces and swells/deswells 
a particle is accepted with probability
\begin{equation}
{\cal P}_{\rm acc}=\min\left\{e^{-\beta(\Delta U_e+\Delta F_p+\Delta E_V+\Delta U_{\rm eff})},~1\right\},
\label{Pacc}
\end{equation}
where $\Delta U_e$, $\Delta F_p$, and $\Delta E_V$ are, respectively, the changes in 
electrostatic self energy [Eq.~(\ref{ue})], polymer free energy [Eq.~(\ref{fp})], and volume energy 
[Eq.~(\ref{EV})] resulting from particle swelling/deswelling, and $\Delta U_{\rm eff}$ is the change in 
internal energy [Eq.~(\ref{Ueff})] associated with elastic and electrostatic interparticle interactions.  
In summing over particle pairs to update $U_{\rm eff}$, we applied the periodic boundary conditions 
to select the image of particle $j$ that is nearest to particle $i$, which amounts to cutting off 
the effective pair potential at a distance $r_c$ equal to half the box length.  For the sizes of
system simulated here, $\kappa r_c\gg 1$, such that finite-size effects are negligible.
In practice, we made simultaneous trial changes in the coordinates $(x, y, z)$ and swelling ratio 
$\alpha$ of each particle with tolerances $\Delta x=\Delta y=\Delta z=0.1 a_0$ and $\Delta\alpha=0.05$.
Through repeated trial moves, the system evolved toward an equilibrium state of minimum total free energy. 

After initializing the particles on the sites of a face-centered-cubic (FCC) lattice, we executed 
a sequence of MC steps, each step consisting of an attempted trial move (displacement and size change)
of every particle.  Following an equilibration stage, after which the total energy fluctuated about a 
stable plateau, we collected statistics by averaging over configurations and computing equilibrium 
thermal and structural properties.  
The intrinsic size polydispersity of the particles was determined by histogramming the swelling ratio 
and computing the probability distribution, $P(\alpha; \phi_0)$, which varies with dry volume fraction.
For structural properties, we computed the radial distribution function $g(r)$, by histogramming 
the center-center separation $r$ between pairs of particles, and the orientationally averaged 
static structure factor from
\begin{equation}
S(q)=1+\frac{2}{N_m}\sum_{i<j=1}^{N_m}\la\frac{\sin(q r_{ij})}{q r_{ij}}\ra,
\label{Sq}
\end{equation}
where $q$ is the scattered wave vector magnitude.  It should be noted that our method, since it
initializes the particles on the sites of a crystal lattice, can determine only whether the 
system is unstable toward melting, but not whether the solid phase is thermodynamically stable.
Identifying equilibrium phase boundaries would require simulating in a different ensemble 
or performing thermodynamic integration to compute total free energies.\cite{frenkel-smit2001}

\subsection{Thermodynamic Perturbation Theory}\label{theory}

To validate our MC simulation method and guide the choice of system parameters, we developed
and implemented a thermodynamic theory based on a variational approximation for the free energy.
Our approach extends to compressible, ionic microgels an approximation previously developed 
and proven accurate for charged colloids\cite{vanRoij1997,denton2006} and for incompressible
ionic microgels.\cite{hedrick-chung-denton2015}  Since size polydispersity associated with swelling
turns out to be minimal, we consider a suspension of microgels all of the same swollen radius.
Combining first-order thermodynamic perturbation theory with a hard-sphere (HS) reference system,
we approximate the constrained excess free energy per microgel for fixed radius $a$: 
\begin{eqnarray}
&&
\hspace*{-0.5cm}
f_{\rm ex}(n_m,a)=\min_{(d)}\left\{f_{\rm HS}(n_m; d)
+2\pi n_m\int_d^{\infty}{\rm d}r\, r^2 g_{\rm HS}(r,n_m; d)
\right. 
\nonumber\\[1ex]
&&\left.{\phantom{\int}}
\times\left[v_H(r,a)+v_{\rm eff}(r,n_m,a)\right]\right\},
\label{fex}
\end{eqnarray}
where $d$ is an effective HS diameter and $f_{\rm HS}$ and $g_{\rm HS}(r)$ are, respectively, 
the excess free energy per particle and radial distribution function of the HS system. 
For a HS fluid, we compute $f_{\rm HS}$ and $g_{\rm HS}(r)$ from the accurate Carnahan-Starling 
and Verlet-Weis expressions.\cite{hansen-mcdonald2006}  
From the Gibbs-Bogoliubov inequality,\cite{hansen-mcdonald2006} minimization with respect to $d$ 
yields a least upper bound to the constrained excess free energy for a given microgel radius.
We note in passing that, in contrast to the case for hard charged colloids, the effective hard-sphere 
diameter for compressible microgels may in principle be smaller than the microgel diameter.
The equilibrium free energy per microgel is finally obtained as the minimum with respect to $a$ 
of the total constrained free energy: 
\begin{equation}
f(n_m)=\min_{(a)}\left\{u_e(a)+f_p(a)+\varepsilon_V(a)+f_{\rm ex}(n_m,a)\right\},
\label{fnm}
\end{equation}
where $\varepsilon_V=E_V/N_m$ is the volume energy per microgel.  The value of $a$ at the minimum
represents the equilibrium swollen microgel radius.  Although we did not compute the osmotic pressure,
we note in passing that this quantity can be computed from the free energy via the thermodynamic relation
\begin{equation}
\pi=n_m^2\left(\frac{\partial f(n_m)}{\partial n_m}\right)_{\frac{N_s}{N_m}},
\label{pressure-theory}
\end{equation}
where the density dependence of the equilibrium particle size must be accounted for in the
derivative.  In the case of Donnan equilibrium, the salt density in the suspension $n_s$ is 
determined by equating the chemical potentials of salt in the suspension and the reservoir:
\begin{equation}
\mu_s=\left(\frac{\partial}{\partial n_s}[n_m(\varepsilon_V+f_m)]\right)_{n_m}=\mu_{sr},
\label{mus}
\end{equation}
explicit expressions for which are given elsewhere.\cite{hedrick-chung-denton2015}

Practical applications of the theory described above are straightforward.  For a given dry volume 
fraction and reservoir salt concentration, numerical implementation involves three nested calculations:
(1) solving Eq.~(\ref{mus}) for $n_s$ via a root-finding algorithm;
(2) minimizing $f(n_m,a)$ with respect to the microgel radius $a$ [Eq.~(\ref{fnm})] via a 
function minimization algorithm; and (3) minimizing $f_{\rm ex}(n_m,a,d)$ with respect to 
the effective hard-sphere diameter $d$ [Eq.~(\ref{fex})].

\section{Results and Discussion}\label{results}

To demonstrate and validate our methods, we studied the dependence of equilibrium particle 
swelling behavior and bulk thermodynamic and structural properties on the concentration of 
ionic microgel suspensions.  For illustration and comparison with previous work, we chose 
the following system parameters, corresponding to deionized aqueous suspensions:
$\lambda_B=0.72$ nm (water at $T=293$ K), $N_m=500$, $Z=500$ or 1000, $a_0=10$ nm, 
$N_{\rm mon}=2\times 10^5$, $N_{\rm ch}=100$, $\chi=0.5$, $B^*=1.5\times 10^4$, $\nu=0.5$, and $n_s=0$. 
From Eq.~(\ref{B2}), these parameters correspond to $Y\simeq 100-1000$ kPa.
In the absence of salt, the screening constant simplifies to $\kappa=\sqrt{4\pi\lambda_B Zn_m}$.
For dry volume fractions $\phi_0$ in the range from 0 to 0.1, we present results computed 
from averages of particle coordinates and radii over 1000 independent configurations, 
separated by intervals of 100 MC steps (total of $10^5$ steps), following an initial 
equilibration stage of $5\times 10^4$ MC steps.

\begin{figure}
\includegraphics[width=0.9\columnwidth]{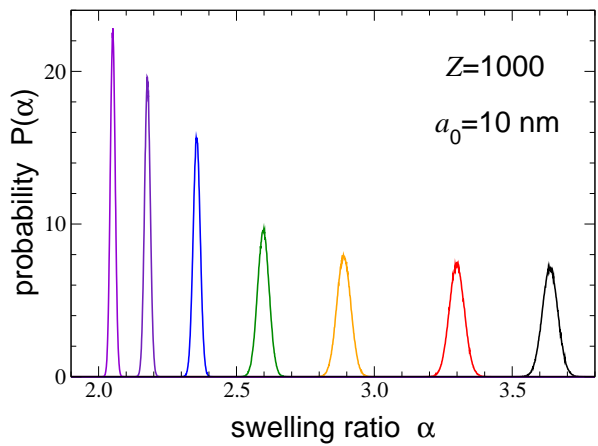}
\vspace*{-0.2cm}
\caption{
Normalized probability distribution $P(\alpha)$ of swelling ratio $\alpha$ in 
deionized suspensions of ionic microgels of valence $Z=1000$ and 
dry radius $a_0=10$ nm, composed of $N_m=2\times 10^5$ monomers with $N_{\rm ch}=100$
chains, in a solvent with Flory solvency parameter $\chi=0.5$ at dry volume fractions 
$\phi_0=0.004$, 0.008, 0.02, 0.04, 0.06, 0.08, 0.1 (right to left).
The particles interact via a Yukawa-Hertz pair potential with reduced Hertz amplitude 
$B^*=1.5\times 10^4$.  With increasing concentration, microgels steadily
deswell, as reflected by the shift from higher to lower swelling ratios. 
}\label{fig3}
\end{figure}

From simulations of the coarse-grained OCM, we computed the probability distribution $P(\alpha)$ 
of the equilibrium swelling ratio in suspensions of ionic microgels of valence $Z=1000$ over a range 
of dry volume fractions.  As shown in Fig.~\ref{fig3}, with increasing concentration, the compressible 
particles progressively deswell, and also become less polydisperse, reflected by $P(\alpha)$ 
shifting to smaller $\alpha$ and narrowing.  Interestingly, the narrowing polydispersity trend 
is opposite that predicted for nonionic microgels.\cite{urich-denton2016}
Note that the low degree of polydispersity seen in the simulations justifies our practical
approximation of equally sized microgels in the thermodynamic perturbation theory.

\begin{figure}
\includegraphics[width=0.9\columnwidth]{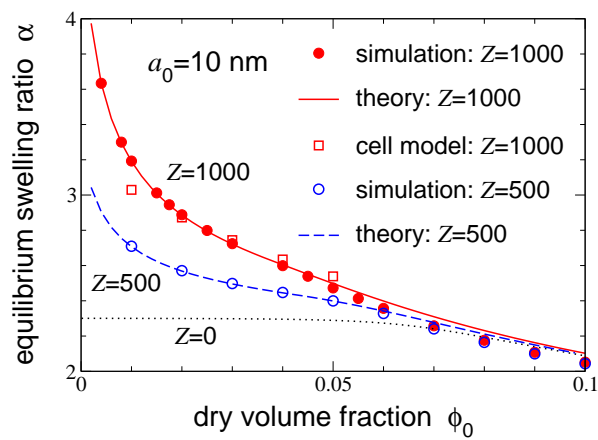}
\vspace*{-0.2cm}
\caption{
Equilibrium swelling ratio $\alpha$ vs.~dry volume fraction $\phi_0$ in deionized suspensions 
of microgels.  Simulation data (circles) are compared with predictions of variational theory (curves) 
and the Poisson-Boltzmann cell model\cite{denton-tang2016} (squares) for the coarse-grained
one-component model of ionic microgels of valence $Z=500$ and 1000.  For comparison, 
theoretical predictions are shown also for nonionic microgels ($Z=0$, dotted curve).  
Other system parameters are the same as in Fig.~\ref{fig3}.  With increasing concentration, 
ionic microgels steadily deswell, while nonionic microgels deswell only above particle overlap 
($\phi_0\simeq 0.06$, $\phi\simeq 0.74$).
}\label{fig4}
\end{figure}

Figure~\ref{fig4} shows both simulation data and theoretical predictions 
for the average equilibrium swelling ratio of both ionic and nonionic microgels vs.~dry volume fraction.
For ionic microgels, the average equilibrium $\alpha$ increases with increasing valence (from $Z=500$ 
to 1000) and exhibits a rapid decrease with $\phi_0$, commencing already in the dilute limit.
In contrast, nonionic microgels ($Z=0$) are much more resistant to deswelling, yielding only 
at concentrations approaching and exceeding particle overlap (around $\phi_0=0.06$).  
It should be noted that for the chain lengths ($N_m/N_{\rm ch}=2000$) and swelling ratios
($\alpha\simeq 2-4$) considered here, the chains are far from fully extended, as required by
the Flory-Rehner approximation for the elastic free energy of network stretching.

Our simulation and perturbation theory implementations of the OCM yield equilibrium swelling ratios 
in near exact agreement at lower concentrations.  Small deviations at higher concentrations, beyond 
particle overlap, may be attributed to approximations inherent to the theory.  In particular, 
the variational approximation gives only a least upper bound to the free energy.  Further, our use 
of the fluid phase expressions for $f_{\rm HS}$ and $g_{\rm HS}(r)$ in Eq.~(\ref{fex}) may forfeit
some accuracy at concentrations where the reference system is actually a solid.  

For comparison, we also show in Fig.~\ref{fig4} previous predictions for the swelling ratio
of ionic microgels computed by applying a newly proposed theorem for the electrostatic component
of swelling.\cite{denton-tang2016}  This theorem -- exact in the spherical cell model -- relates 
the electrostatic contribution to the osmotic pressure across the periphery of a permeable macroion 
to the microion density profiles, which we computed by solving the nonlinear Poisson-Boltzmann (PB) 
equation in the cell model.  Good agreement between, on the one hand, our simulation and linear 
theory implementations of the OCM and, on the other hand, the nonlinear PB theory implementation 
of the cell model provides an important validation of the new osmotic pressure theorem and also 
justifies the linear response approximation.  Relatively small deviations at the lowest and highest 
concentrations may be attributed to differences between the OCM and the cell model and weak 
nonlinear screening effects.

In previous work,\cite{denton-tang2016} we also performed molecular dynamics simulations of
ionic microgels in the cell model with explicit counterions.  Close agreement between the
simulation data and predictions of PB theory for counterion density profiles and microgel
swelling ratios validate the PB theory, confirming that correlations between monovalent 
counterions are weak and can be neglected in the systems considered here.

\begin{figure}
\includegraphics[width=0.9\columnwidth]{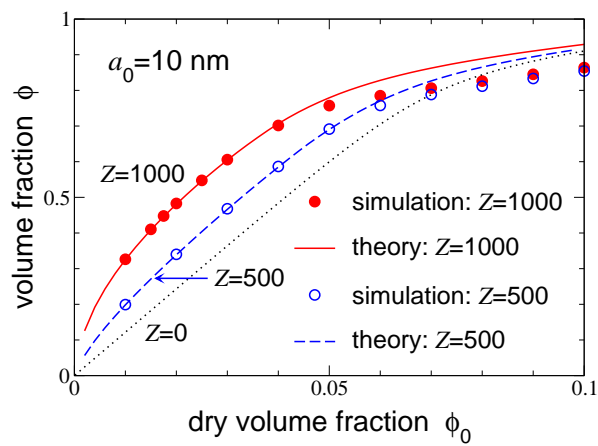}
\vspace*{-0.2cm}
\caption{
Equilibrium volume fraction $\phi$ vs.~dry volume fraction $\phi_0$ in deionized suspensions of 
microgels.  Simulation data (symbols) are compared with predictions of variational theory (curves)
for the coarse-grained one-component model of ionic microgels of valence $Z=500$ and 1000.  
For comparison, theoretical predictions are shown also for nonionic microgels ($Z=0$, dotted curve).  
Other system parameters are same as in Fig.~\ref{fig3}.
}\label{fig5}
\end{figure}
\begin{figure}[h!]
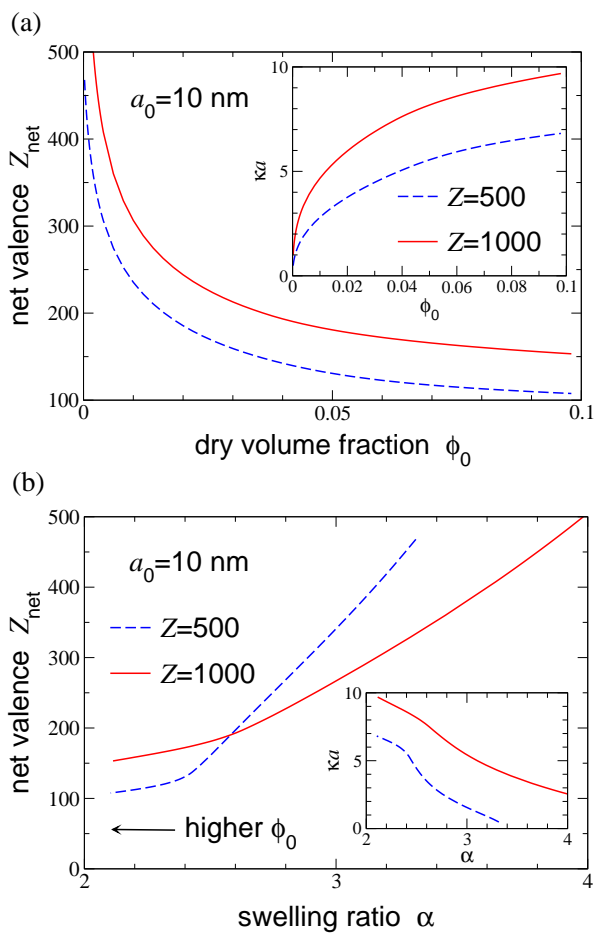

\includegraphics[width=0.9\columnwidth]{znet-vs-phi0.a10.inset.eps}
\includegraphics[width=0.9\columnwidth]{znet-vs-alpha.a10.inset.eps}
\vspace*{-0.2cm}
\caption{
Theoretical predictions [from Eq.~(\ref{znet})] for net valence $Z_{\rm net}$ of ionic microgels 
(valences $Z=500$ and 1000) in bulk suspensions, for same system parameters as in Fig.~\ref{fig3},
as a function of (a) dry volume fraction and (b) swelling ratio.  
Insets show product of screening constant $\kappa$ and swollen radius $a$.
}\label{fig6}
\end{figure}

In a complementary illustration of particle deswelling, Fig.~\ref{fig5} displays the variation
of actual volume fraction with dry volume fraction.  For nonionic microgels, $\phi$ is simply
proportional to $\phi_0$ at lower concentrations, with nonlinear dependence developing only at 
concentrations exceeding particle overlap, where elastic (Hertz) interactions become significant.
In sharp contrast, ionic microgels are considerably more swollen by their electrostatic self energy 
and fill a volume fraction that varies nonlinearly with respect to $\phi_0$ -- with negative curvature 
-- over the whole concentration range.  This sensitive dependence on concentration results from a 
complex interplay between single-particle free energy and effective electrostatic interactions, 
including the volume energy and relatively long-range Yukawa pair interactions.  Note that any 
discrepancies between simulation and theory are amplified by the cubic dependence of $\phi$ on $\alpha$.

As ionic microgels swell or deswell, the numbers of counterions inside and outside the particles
can vary, thus affecting the net valence $Z_{\rm net}$.  Figure~\ref{fig6} presents our predictions 
for $Z_{\rm net}$, computed from Eq.~(\ref{znet}), for microgels in a bulk suspension with the same 
system parameters as in Figs.~\ref{fig3}-\ref{fig5}.  
To interpret these results, it is important to bear in mind that, as $\phi_0$ and  $\alpha$ vary,
the screening constant $\kappa$ also varies, as shown in the insets to Fig.~\ref{fig6}.  
With increasing concentration, as the particles deswell, the fraction of interior counterions rises, 
thus reducing $Z_{\rm net}$.  Considered as a function of swelling ratio, however, $Z_{\rm net}$ 
increases with $\alpha$, since $\kappa a$ decreases with $\alpha$.  

For the systems considered here, $Z_{\rm net}\lambda_B/a$, a measure of electrostatic 
coupling strength, ranges from 5-10. The close agreement between swelling predictions 
from the linearized PB theory implementation of the OCM and the nonlinear PB theory 
implementation of the cell model (Fig.~\ref{fig4}) suggests that nonlinear screening effects 
are weak here.  At stronger couplings, however, such that $Z_{\rm net}\lambda_B/a>\mathcal{O}(10)$, 
nonlinear effects may become significant for ionic microgels, as demonstrated in two recent studies.
\cite{braibanti-perez2016,quesada-perez-moncho-jorda2018} 
In such cases, the linearized theory can be extended into the nonlinear regime via 
charge renormalization schemes that incorporate an effective macroion charge.\cite{colla-likos2014}
We note in passing that the swollen microgels considered here are substantially (3-5 times) larger 
than those modeled by Quesada-P\'erez \etalia\cite{quesada-perez-moncho-jorda2018} 

\begin{figure}[t!]
\includegraphics[width=0.9\columnwidth]{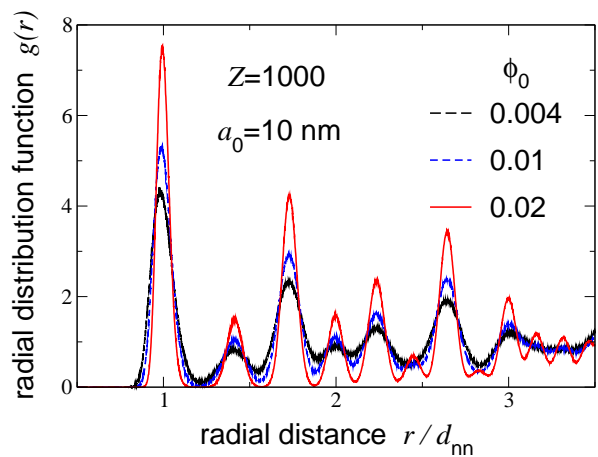}
\vspace*{-0.2cm}
\caption{
Radial distribution function $g(r)$ vs.~radial distance $r$, in units of 
nearest-neighbor distance $d_{\rm nn}$ in FCC lattice, in suspensions of ionic,
compressible microgels with same system parameters as in Fig.~\ref{fig3}.
Results are shown for dry volume fractions $\phi_0=0.004$ (dashed black curve),
0.01 (short-dashed blue curve), and 0.02 (solid red curve).  These systems
are all in an FCC crystal phase, as revealed by the positions of the peaks.
}\label{fig7}
\end{figure}
\begin{figure}
\includegraphics[width=0.9\columnwidth]{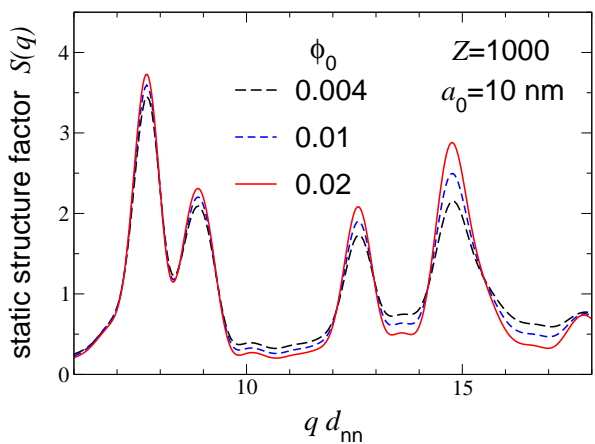}
\vspace*{-0.2cm}
\caption{
Static structure factor $S(q)$ [from Eq.~(\ref{Sq})] vs.~scattered wave vector magnitude $q$, 
in units of inverse nearest-neighbor distance $d_{\rm nn}$ in FCC lattice.  Results are shown 
for dry volume fractions $\phi_0=0.004$ (dashed black curve), 0.01 (short-dashed blue curve), 
and 0.02 (solid red curve), corresponding to radial distribution functions in Fig.~\ref{fig7}.  
These suspensions are all in an FCC crystal phase, as reflected by the height of the main peak,
$S(q_{\rm max})>2.85$.
}\label{fig8}
\end{figure}
\begin{figure}[t!]
\includegraphics[width=0.9\columnwidth]{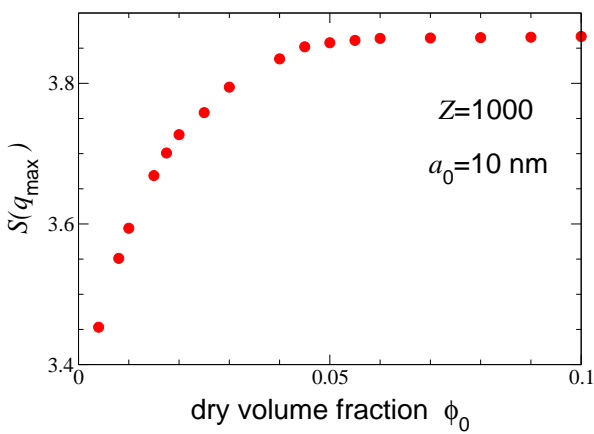}
\vspace*{-0.2cm}
\caption{
Main peak height of static structure factor $S(q_{\rm max})$ vs.~dry volume fraction $\phi_0$
for same systems as represented in Figs.~\ref{fig7} and \ref{fig8}.  With increasing concentration,
pair correlations strengthen until, near particle overlap ($\phi_0\simeq 0.06$), the structure saturates.
}\label{fig9}
\end{figure}

To quantify the variation of bulk structure with concentration, we computed radial distribution 
functions and static structure factors.  Figures~\ref{fig7} and ~\ref{fig8} show our simulation 
data for $g(r)$ and $S(q)$, respectively, for a series of dry volume fractions, illustrating 
the strengthening of correlations between microgels with increasing concentration.  
The positions and heights of the distinct peaks indicate crystalline ordering and reveal that 
the system has retained its initial FCC structure.  The height of the main peak of $S(q)$
exceeds the threshold of 2.85 set by the Hansen-Verlet freezing criterion,\cite{hansen-verlet1969}
suggesting that the system is in a stable solid phase.  As seen in Fig.~\ref{fig9}, the main peak 
height grows steadily with increasing concentration.  Interestingly, however, $S(q_{\rm max})$ 
plateaus beyond particle overlap.  This unusual structural saturation seems to indicate that the 
soft particles are free to wander around their equilibrium sites, even in a dense crystal structure.

To further test the coarse-grained OCM, we compare with recent light scattering measurements of the 
equilibrium swollen sizes of loosely cross-linked PNIPAM-co-PAA microgels in deionized aqueous suspensions.
\cite{holmqvist-shurtenberger2012}  
Setting the bare valence and dry radius at their respective measured values 
of $Z=3.5\times 10^4$ and $a_0=50$ nm ($N_{\rm mon}=3\times 10^6$),
and treating the cross-linker fraction, $x\equiv N_{\rm ch}/N_{\rm mon}$, and Flory $\chi$ 
parameter as fitting parameters, we computed swelling ratios and compared predictions with 
experimental data (using corrected concentrations\cite{holmqvist-shurtenberger2012-erratum}: 
0.0053, 0.0060, 0.0100, 0.0167, 0.053 $\mu$M).

\begin{figure}
\includegraphics[width=0.9\columnwidth]{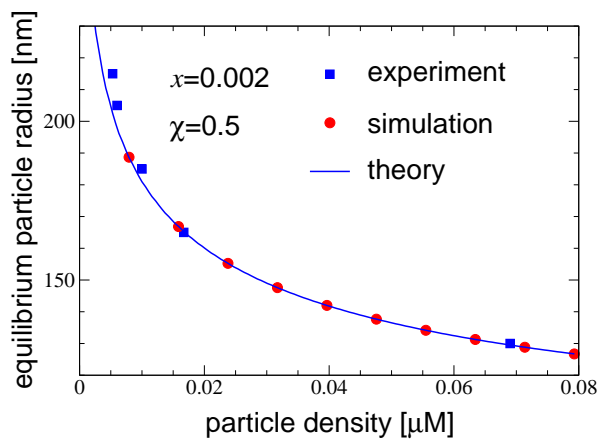}
\vspace*{-0.2cm}
\caption{
Equilibrium particle radius vs.~particle density for ionic microgels.
Simulation data (red circles) and predictions of variational theory for one-component model (curve) 
are compared with experimental data\cite{holmqvist-shurtenberger2012,holmqvist-shurtenberger2012-erratum} 
(blue squares) for PNIPAM-co-PAA microgels in deionized aqueous suspensions with system parameters
$Z=3.5\times 10^4$ and $a_0=50$ nm ($N_{\rm mon}=3\times 10^6$).  Fitting parameters are 
cross-linker fraction $x=N_{\rm ch}/N_{\rm mon}$ and Flory solvency parameter $\chi$.
}\label{fig10}
\end{figure}

As seen in Fig.~\ref{fig10}, theory and simulation are in near-perfect 
agreement and the OCM accurately fits the data at least as well as the cell model, which neglects 
elastic interparticle interactions.\cite{denton-tang2016}  It should be noted, however, that the 
best-fit values of the free parameters may be somewhat unphysical.  The fitted value of $x$ is 
likely lower than the actual average cross-linker fraction, which may reflect differences between 
the assumed homogeneous cross-linker distribution and the actual core-shell structure of the particles.
The fitted value of $\chi$ is likely higher than the actual value for PNIPAM in water, which would 
be consistent with the recently documented tendency of the Flory-Rehner theory to overestimate $\chi$.
\cite{richtering2017}  Increasing $x$ or $\chi$ would lower the predicted swelling ratio.
Despite these potential limitations, the model at least qualitatively explains the concentration 
dependence of ionic microgel swelling.

\begin{figure}[t!]
\includegraphics[width=0.9\columnwidth]{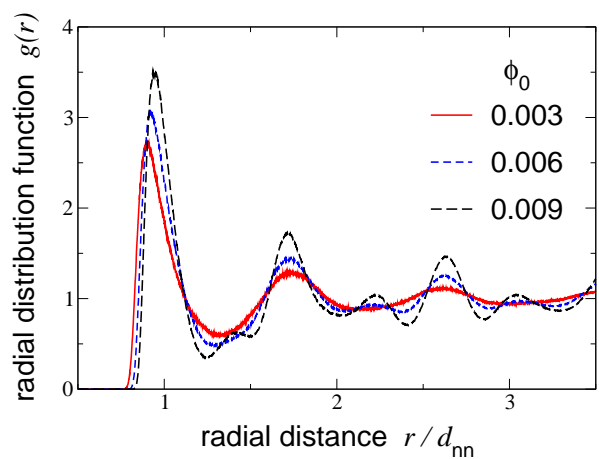}
\vspace*{-0.2cm}
\caption{
Radial distribution function $g(r)$ vs.~radial distance $r$, in units of 
nearest-neighbor distance $d_{\rm nn}$ in FCC lattice, in suspensions of ionic,
compressible microgels with same system parameters as in Fig.~\ref{fig10}.
Results are shown for dry volume fractions $\phi_0=0.003$ (solid red curve),
which is in a fluid phase, and $\phi_0=0.006$ (short-dashed blue curve) 
and 0.009 (dashed black curve), both of which are in FCC crystal phases,
as revealed by positions of peaks.
}\label{fig11}
\end{figure}

\begin{figure}
\includegraphics[width=0.9\columnwidth]{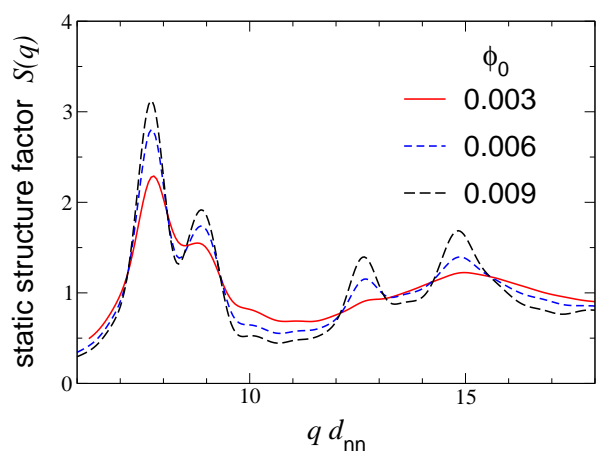}
\vspace*{-0.2cm}
\caption{
Static structure factor $S(q)$ vs.~scattered wave vector magnitude $q$, 
in units of inverse nearest-neighbor distance $d_{\rm nn}$ in FCC lattice.
Results are shown for dry volume fractions $\phi_0=0.003$ (solid red curve),
$\phi_0=0.006$ (short-dashed blue curve), and 0.009 (dashed black curve),
corresponding to radial distribution functions in Fig.~\ref{fig11}.
}\label{fig12}
\end{figure}

Beyond the swelling ratio, our simulations also yield the radial distribution function 
(Fig.~\ref{fig11}) and static structure factor (Fig.~\ref{fig12}).  With increasing
concentration, the peaks of $g(r)$ and $S(q)$ grow taller and more distinct, reflecting 
strengthening correlations between microgels.  From the heights of the main peaks of $S(q)$,
our results suggest that the systems with dry volume fractions $\phi_0=0.003$, 0.006, 
and 0.009 are, respectively, in a disordered fluid phase, on the verge of freezing, 
and in a solid phase with crystalline order.  We emphasize again, however, that our simulation method 
cannot distinguish between stable and metastable solid states.

Finally, it should be noted that, although we have here applied the coarse-grained model only to
salt-free (deionized) microgel suspensions, our approach can be easily applied to salty suspensions
with implicit salt ions.  For a closed system, the density of salt ions must simply be included 
in the effective electrostatic interactions, as described in Sec.~\ref{coarse-grained-model}.  
For a system in Donnan equilibrium with a salt reservoir, the salt concentration in the system 
must first be determined by equating the chemical potentials of salt in the system and reservoir.
In previous work,\cite{denton-tang2016} we showed that, for the system parameters of Fig.~\ref{fig3}, 
the linear-response theory implementation of the cell model predicts monotonic deswelling of 
ionic microgels with increasing salt concentration.  We expect the OCM to yield similar 
predictions for the dependence of $\alpha$ on salt concentration.

\section{Summary and Conclusions}\label{conclusions}

In summary, we developed a Monte Carlo simulation algorithm and a thermodynamic perturbation theory 
for a coarse-grained model of compressible, ionic microgel suspensions and studied the concentration 
dependence of bulk thermal and structural properties.  The model incorporates both the colloidal 
and the polymeric natures of ionic gel particles into an effective Hamiltonian comprising 
one- and two-body effective elastic and electrostatic interactions.  As far as we are aware, our model 
is the first to consistently account for both elastic and electrostatic influences on the swelling of 
ionic microgels in a bulk suspension.  

As an illustrative application, we investigated equilibrium particle swelling and structure of bulk 
suspensions with selected system parameters.  Specifically, we computed equilibrium particle size 
distributions, swelling ratios, volume fractions, net valences, radial distribution functions, and 
static structure factors.  Close agreement between swelling ratios independently computed from theory 
and simulation validated our computational methods.  With increasing concentration, swelling ratios 
of ionic microgels decrease more precipitously than those of nonionic microgels, while net valences 
also decrease monotonically.  The simulations further revealed an unusual saturation of pair correlations 
with increasing concentration beyond particle overlap.

To further test our methods, we compared the predicted swelling behavior against experimental data 
for deionized, aqueous suspensions of PNIPAM microgels.  Close agreement between simulation, theory,
and experiment supports the predictive power of our approach.  The coarse-grained model and methods 
developed and demonstrated here provide a reasonably accurate and computationally efficient path to 
modeling swelling and structural properties of bulk suspensions of ionic microgels.  The predictions
may help to guide and interpret future experiments and may clarify the importance of including 
particle swelling in modeling ionic microgel suspensions.  

The coarse-grained model could be refined by incorporating a more accurate theory of the 
single-particle polymer network free energy than provided by the Flory-Rehner theory and by 
improving upon the Hertz theory of elastic pair interactions.
Furthermore, the model can be extended to describe microgels with inhomogeneous distributions of 
cross-linkers and fixed charges.  Future work will include computing bulk osmotic pressure 
and phase behavior, which will require consistently accounting for concentration dependence of 
the single-particle free energy and effective interparticle interactions, and incorporating
charge renormalization schemes to model more highly charged microgels.

\vspace*{0.5cm}
\noindent{\bf \large Acknowledgments} \\[1ex]
This paper is dedicated to the memory of Per Linse, whose many important insights into soft matter
and electrostatics will have lasting impact, and whose generous spirit is deeply missed.  
Helpful discussions with Jan Dhont, Gerhard N\"agele, Mariano Brito, and Peter Schurtenberger 
are gratefully acknowledged.
Parts of this work were supported by the National Science Foundation (Grant No.~DMR-1106331).

\balance

\bibliographystyle{rsc}

\begin{mcitethebibliography}{83}
\providecommand*{\natexlab}[1]{#1}
\providecommand*{\mciteSetBstSublistMode}[1]{}
\providecommand*{\mciteSetBstMaxWidthForm}[2]{}
\providecommand*{\mciteBstWouldAddEndPuncttrue}
  {\def\EndOfBibitem{\unskip.}}
\providecommand*{\mciteBstWouldAddEndPunctfalse}
  {\let\EndOfBibitem\relax}
\providecommand*{\mciteSetBstMidEndSepPunct}[3]{}
\providecommand*{\mciteSetBstSublistLabelBeginEnd}[3]{}
\providecommand*{\EndOfBibitem}{}
\mciteSetBstSublistMode{f}
\mciteSetBstMaxWidthForm{subitem}
{(\emph{\alph{mcitesubitemcount}})}
\mciteSetBstSublistLabelBeginEnd{\mcitemaxwidthsubitemform\space}
{\relax}{\relax}

\bibitem[Baker(1949)]{baker1949}
W.~O. Baker, \emph{Ind. Eng. Chem.}, 1949, \textbf{41}, 511--520\relax
\mciteBstWouldAddEndPuncttrue
\mciteSetBstMidEndSepPunct{\mcitedefaultmidpunct}
{\mcitedefaultendpunct}{\mcitedefaultseppunct}\relax
\EndOfBibitem
\bibitem[Pelton and Chibante(1986)]{pelton1986}
R.~H. Pelton and P.~Chibante, \emph{Colloids Surf.}, 1986, \textbf{20},
  247--256\relax
\mciteBstWouldAddEndPuncttrue
\mciteSetBstMidEndSepPunct{\mcitedefaultmidpunct}
{\mcitedefaultendpunct}{\mcitedefaultseppunct}\relax
\EndOfBibitem
\bibitem[Pelton(2000)]{pelton2000}
R.~H. Pelton, \emph{Adv. Colloid Interface Sci.}, 2000, \textbf{85},
  1--33\relax
\mciteBstWouldAddEndPuncttrue
\mciteSetBstMidEndSepPunct{\mcitedefaultmidpunct}
{\mcitedefaultendpunct}{\mcitedefaultseppunct}\relax
\EndOfBibitem
\bibitem[Saunders \emph{et~al.}(2009)Saunders, Laajam, Daly, Teow, Hu, and
  Stepto]{saunders2009}
B.~R. Saunders, N.~Laajam, E.~Daly, S.~Teow, X.~Hu and R.~Stepto, \emph{Adv.
  Colloid Interface Sci.}, 2009, \textbf{147}, 251--262\relax
\mciteBstWouldAddEndPuncttrue
\mciteSetBstMidEndSepPunct{\mcitedefaultmidpunct}
{\mcitedefaultendpunct}{\mcitedefaultseppunct}\relax
\EndOfBibitem
\bibitem[Lyon and Serpe(2012)]{HydrogelBook2012}
\emph{Hydrogel Micro and Nanoparticles}, ed. L.~A. Lyon and M.~J. Serpe,
  Wiley-VCH Verlag GmbH \& Co. KGaA, Weinheim, 2012\relax
\mciteBstWouldAddEndPuncttrue
\mciteSetBstMidEndSepPunct{\mcitedefaultmidpunct}
{\mcitedefaultendpunct}{\mcitedefaultseppunct}\relax
\EndOfBibitem
\bibitem[Fern{\'a}ndez-Nieves \emph{et~al.}(2011)Fern{\'a}ndez-Nieves, Wyss,
  Mattsson, and Weitz]{MicrogelBook2011}
\emph{Microgel Suspensions: Fundamentals and Applications}, ed.
  A.~Fern{\'a}ndez-Nieves, H.~Wyss, J.~Mattsson and D.~A. Weitz, Wiley-VCH
  Verlag GmbH \& Co. KGaA, Weinheim, 2011\relax
\mciteBstWouldAddEndPuncttrue
\mciteSetBstMidEndSepPunct{\mcitedefaultmidpunct}
{\mcitedefaultendpunct}{\mcitedefaultseppunct}\relax
\EndOfBibitem
\bibitem[Lyon and Fern{\'a}ndez-Nieves(2012)]{lyon-nieves-AnnuRevPhysChem2012}
L.~A. Lyon and A.~Fern{\'a}ndez-Nieves, \emph{Annu. Rev. Phys. Chem.}, 2012,
  \textbf{63}, 25--43\relax
\mciteBstWouldAddEndPuncttrue
\mciteSetBstMidEndSepPunct{\mcitedefaultmidpunct}
{\mcitedefaultendpunct}{\mcitedefaultseppunct}\relax
\EndOfBibitem
\bibitem[Yunker \emph{et~al.}(2014)Yunker, Chen, Gratale, Lohr, Still, and
  Yodh]{yunker-yodh-review2014}
P.~J. Yunker, K.~Chen, D.~Gratale, M.~A. Lohr, T.~Still and A.~G. Yodh,
  \emph{Rep. Prog. Phys.}, 2014, \textbf{77}, 056601--1--29\relax
\mciteBstWouldAddEndPuncttrue
\mciteSetBstMidEndSepPunct{\mcitedefaultmidpunct}
{\mcitedefaultendpunct}{\mcitedefaultseppunct}\relax
\EndOfBibitem
\bibitem[Borrega \emph{et~al.}(1999)Borrega, Cloitre, Betremieux, Ernst, and
  Leibler]{cloitre-leibler1999}
R.~Borrega, M.~Cloitre, I.~Betremieux, B.~Ernst and L.~Leibler, \emph{Euro.
  Phys. Lett.}, 1999, \textbf{47}, 729--735\relax
\mciteBstWouldAddEndPuncttrue
\mciteSetBstMidEndSepPunct{\mcitedefaultmidpunct}
{\mcitedefaultendpunct}{\mcitedefaultseppunct}\relax
\EndOfBibitem
\bibitem[Cloitre \emph{et~al.}(2003)Cloitre, Borrega, Monti, and
  Leibler]{cloitre-leibler2003}
M.~Cloitre, R.~Borrega, F.~Monti and L.~Leibler, \emph{C. R. Physique}, 2003,
  \textbf{4}, 221--230\relax
\mciteBstWouldAddEndPuncttrue
\mciteSetBstMidEndSepPunct{\mcitedefaultmidpunct}
{\mcitedefaultendpunct}{\mcitedefaultseppunct}\relax
\EndOfBibitem
\bibitem[Tan \emph{et~al.}(2004)Tan, Tam, Lam, and Tan]{tan2004}
B.~H. Tan, K.~C. Tam, Y.~C. Lam and C.~B. Tan, \emph{J. Rheol.}, 2004,
  \textbf{48}, 915--926\relax
\mciteBstWouldAddEndPuncttrue
\mciteSetBstMidEndSepPunct{\mcitedefaultmidpunct}
{\mcitedefaultendpunct}{\mcitedefaultseppunct}\relax
\EndOfBibitem
\bibitem[Fern{\'a}ndez-Nieves \emph{et~al.}(2000)Fern{\'a}ndez-Nieves,
  Fern{\'a}ndez-Barbero, Vincent, and de~las {Nieves}]{nieves-macromol2000}
A.~Fern{\'a}ndez-Nieves, A.~Fern{\'a}ndez-Barbero, B.~Vincent and F.~J. de~las
  {Nieves}, \emph{Macromol.}, 2000, \textbf{33}, 2114--2118\relax
\mciteBstWouldAddEndPuncttrue
\mciteSetBstMidEndSepPunct{\mcitedefaultmidpunct}
{\mcitedefaultendpunct}{\mcitedefaultseppunct}\relax
\EndOfBibitem
\bibitem[Fern{\'a}ndez-Nieves \emph{et~al.}(2003)Fern{\'a}ndez-Nieves,
  Fern{\'a}ndez-Barbero, Vincent, and de~las {Nieves}]{nieves-jcp2003}
A.~Fern{\'a}ndez-Nieves, A.~Fern{\'a}ndez-Barbero, B.~Vincent and F.~J. de~las
  {Nieves}, \emph{J. Chem. Phys.}, 2003, \textbf{119}, 10383--10388\relax
\mciteBstWouldAddEndPuncttrue
\mciteSetBstMidEndSepPunct{\mcitedefaultmidpunct}
{\mcitedefaultendpunct}{\mcitedefaultseppunct}\relax
\EndOfBibitem
\bibitem[Li{\'e}tor-Santos \emph{et~al.}(2009)Li{\'e}tor-Santos,
  Sierra-Mart{\'i}n, Vavrin, Hu, Gasser, and
  Fern{\'a}ndez-Nieves]{nieves-macromol2009}
J.~J. Li{\'e}tor-Santos, B.~Sierra-Mart{\'i}n, R.~Vavrin, Z.~Hu, U.~Gasser and
  A.~Fern{\'a}ndez-Nieves, \emph{Macromol.}, 2009, \textbf{42},
  6225--6230\relax
\mciteBstWouldAddEndPuncttrue
\mciteSetBstMidEndSepPunct{\mcitedefaultmidpunct}
{\mcitedefaultendpunct}{\mcitedefaultseppunct}\relax
\EndOfBibitem
\bibitem[Hertle \emph{et~al.}(2010)Hertle, Zeiser, Hasen{\"o}hrl, Busch, and
  Hellweg]{hellweg2010}
Y.~Hertle, M.~Zeiser, C.~Hasen{\"o}hrl, P.~Busch and T.~Hellweg, \emph{Colloid
  Polym. Sci.}, 2010, \textbf{288}, 1047--1059\relax
\mciteBstWouldAddEndPuncttrue
\mciteSetBstMidEndSepPunct{\mcitedefaultmidpunct}
{\mcitedefaultendpunct}{\mcitedefaultseppunct}\relax
\EndOfBibitem
\bibitem[Menut \emph{et~al.}(2012)Menut, Seiffert, Sprakel, and
  Weitz]{weitz-sm2012}
P.~Menut, S.~Seiffert, J.~Sprakel and D.~A. Weitz, \emph{Soft Matter}, 2012,
  \textbf{8}, 156--164\relax
\mciteBstWouldAddEndPuncttrue
\mciteSetBstMidEndSepPunct{\mcitedefaultmidpunct}
{\mcitedefaultendpunct}{\mcitedefaultseppunct}\relax
\EndOfBibitem
\bibitem[Romeo \emph{et~al.}(2012)Romeo, Imperiali, Kim, Fern{\'a}ndez-Nieves,
  and Weitz]{weitz-jcp2012}
G.~Romeo, L.~Imperiali, J.-W. Kim, A.~Fern{\'a}ndez-Nieves and D.~A. Weitz,
  \emph{J. Chem. Phys.}, 2012, \textbf{136}, 124905--1--9\relax
\mciteBstWouldAddEndPuncttrue
\mciteSetBstMidEndSepPunct{\mcitedefaultmidpunct}
{\mcitedefaultendpunct}{\mcitedefaultseppunct}\relax
\EndOfBibitem
\bibitem[Romeo and Ciamarra(2013)]{ciamarra2013}
G.~Romeo and M.~P. Ciamarra, \emph{Soft Matter}, 2013, \textbf{9},
  5401--5406\relax
\mciteBstWouldAddEndPuncttrue
\mciteSetBstMidEndSepPunct{\mcitedefaultmidpunct}
{\mcitedefaultendpunct}{\mcitedefaultseppunct}\relax
\EndOfBibitem
\bibitem[Li{\'e}tor-Santos \emph{et~al.}(2011)Li{\'e}tor-Santos,
  Sierra-Mart{\'i}n, Gasser, and Fern{\'a}ndez-Nieves]{nieves-sm2011}
J.~J. Li{\'e}tor-Santos, B.~Sierra-Mart{\'i}n, U.~Gasser and
  A.~Fern{\'a}ndez-Nieves, \emph{Soft Matter}, 2011, \textbf{7},
  6370--6374\relax
\mciteBstWouldAddEndPuncttrue
\mciteSetBstMidEndSepPunct{\mcitedefaultmidpunct}
{\mcitedefaultendpunct}{\mcitedefaultseppunct}\relax
\EndOfBibitem
\bibitem[Riest \emph{et~al.}(2012)Riest, Mohanty, Schurtenberger, and
  Likos]{schurtenberger-ZPC2012}
J.~Riest, P.~Mohanty, P.~Schurtenberger and C.~N. Likos, \emph{Z. Phys. Chem.},
  2012, \textbf{226}, 711--735\relax
\mciteBstWouldAddEndPuncttrue
\mciteSetBstMidEndSepPunct{\mcitedefaultmidpunct}
{\mcitedefaultendpunct}{\mcitedefaultseppunct}\relax
\EndOfBibitem
\bibitem[Sierra-Mart{\'i}n and Fern{\'a}ndez-Nieves(2012)]{nieves-sm2012}
B.~Sierra-Mart{\'i}n and A.~Fern{\'a}ndez-Nieves, \emph{Soft Matter}, 2012,
  \textbf{8}, 4141--4150\relax
\mciteBstWouldAddEndPuncttrue
\mciteSetBstMidEndSepPunct{\mcitedefaultmidpunct}
{\mcitedefaultendpunct}{\mcitedefaultseppunct}\relax
\EndOfBibitem
\bibitem[Li{\'e}tor-Santos \emph{et~al.}(2011)Li{\'e}tor-Santos,
  Sierra-Mart{\'i}n, and Fern{\'a}ndez-Nieves]{nieves-bulk-shear-pre2011}
J.~J. Li{\'e}tor-Santos, B.~Sierra-Mart{\'i}n and A.~Fern{\'a}ndez-Nieves,
  \emph{Phys. Rev. E}, 2011, \textbf{84}, 060402(R)--1--4\relax
\mciteBstWouldAddEndPuncttrue
\mciteSetBstMidEndSepPunct{\mcitedefaultmidpunct}
{\mcitedefaultendpunct}{\mcitedefaultseppunct}\relax
\EndOfBibitem
\bibitem[Sierra-Mart{\'i}n \emph{et~al.}(2011)Sierra-Mart{\'i}n, Laporte,
  South, Lyon, and Fern{\'a}ndez-Nieves]{nieves-bulk-pre2011}
B.~Sierra-Mart{\'i}n, Y.~Laporte, A.~B. South, L.~A. Lyon and
  A.~Fern{\'a}ndez-Nieves, \emph{Phys. Rev. E}, 2011, \textbf{84},
  011406--1--4\relax
\mciteBstWouldAddEndPuncttrue
\mciteSetBstMidEndSepPunct{\mcitedefaultmidpunct}
{\mcitedefaultendpunct}{\mcitedefaultseppunct}\relax
\EndOfBibitem
\bibitem[Hashmi and Dufresne(2009)]{dufresne2009}
S.~M. Hashmi and E.~R. Dufresne, \emph{Soft Matter}, 2009, \textbf{5},
  3682--3688\relax
\mciteBstWouldAddEndPuncttrue
\mciteSetBstMidEndSepPunct{\mcitedefaultmidpunct}
{\mcitedefaultendpunct}{\mcitedefaultseppunct}\relax
\EndOfBibitem
\bibitem[Pelaez-Fernandez \emph{et~al.}(2015)Pelaez-Fernandez, Souslov, Lyon,
  Goldbart, and Fern{\'a}ndez-Nieves]{nieves-prl2015}
M.~Pelaez-Fernandez, A.~Souslov, L.~A. Lyon, P.~M. Goldbart and
  A.~Fern{\'a}ndez-Nieves, \emph{\PRL}, 2015, \textbf{114}, 098303--1--5\relax
\mciteBstWouldAddEndPuncttrue
\mciteSetBstMidEndSepPunct{\mcitedefaultmidpunct}
{\mcitedefaultendpunct}{\mcitedefaultseppunct}\relax
\EndOfBibitem
\bibitem[Mohanty and Richtering(2008)]{mohanty-richtering2008}
P.~S. Mohanty and W.~Richtering, \emph{J. Phys. Chem. B}, 2008, \textbf{112},
  14692--14697\relax
\mciteBstWouldAddEndPuncttrue
\mciteSetBstMidEndSepPunct{\mcitedefaultmidpunct}
{\mcitedefaultendpunct}{\mcitedefaultseppunct}\relax
\EndOfBibitem
\bibitem[Eckert and Richter(2008)]{richtering2008}
T.~Eckert and W.~Richter, \emph{J. Chem. Phys.}, 2008, \textbf{129},
  124902--1--6\relax
\mciteBstWouldAddEndPuncttrue
\mciteSetBstMidEndSepPunct{\mcitedefaultmidpunct}
{\mcitedefaultendpunct}{\mcitedefaultseppunct}\relax
\EndOfBibitem
\bibitem[St.~John \emph{et~al.}(2007)St.~John, Breedveld, and Lyon]{lyon2007}
A.~N. St.~John, V.~Breedveld and L.~A. Lyon, \emph{J. Phys. Chem. B}, 2007,
  \textbf{111}, 7796--7801\relax
\mciteBstWouldAddEndPuncttrue
\mciteSetBstMidEndSepPunct{\mcitedefaultmidpunct}
{\mcitedefaultendpunct}{\mcitedefaultseppunct}\relax
\EndOfBibitem
\bibitem[Muluneh and Weitz(2012)]{weitz-pre2012}
M.~Muluneh and D.~A. Weitz, \emph{Phys. Rev. E}, 2012, \textbf{85},
  021405--1--6\relax
\mciteBstWouldAddEndPuncttrue
\mciteSetBstMidEndSepPunct{\mcitedefaultmidpunct}
{\mcitedefaultendpunct}{\mcitedefaultseppunct}\relax
\EndOfBibitem
\bibitem[Mohanty \emph{et~al.}(2012)Mohanty, Yethiraj, and
  Schurtenberger]{schurtenberger-SM2012}
P.~S. Mohanty, A.~Yethiraj and P.~Schurtenberger, \emph{Soft Matter}, 2012,
  \textbf{8}, 10819--10822\relax
\mciteBstWouldAddEndPuncttrue
\mciteSetBstMidEndSepPunct{\mcitedefaultmidpunct}
{\mcitedefaultendpunct}{\mcitedefaultseppunct}\relax
\EndOfBibitem
\bibitem[Holmqvist \emph{et~al.}(2012)Holmqvist, Mohanty, N{\"a}gele,
  Schurtenberger, and Heinen]{holmqvist-shurtenberger2012}
P.~Holmqvist, P.~S. Mohanty, G.~N{\"a}gele, P.~Schurtenberger and M.~Heinen,
  \emph{\PRL}, 2012, \textbf{109}, 048302--1--5\relax
\mciteBstWouldAddEndPuncttrue
\mciteSetBstMidEndSepPunct{\mcitedefaultmidpunct}
{\mcitedefaultendpunct}{\mcitedefaultseppunct}\relax
\EndOfBibitem
\bibitem[Holmqvist \emph{et~al.}(2016)Holmqvist, Mohanty, N{\"a}gele,
  Schurtenberger, and Heinen]{holmqvist-shurtenberger2012-erratum}
P.~Holmqvist, P.~S. Mohanty, G.~N{\"a}gele, P.~Schurtenberger and M.~Heinen,
  \emph{\PRL}, 2016, \textbf{117}, 179901(E)\relax
\mciteBstWouldAddEndPuncttrue
\mciteSetBstMidEndSepPunct{\mcitedefaultmidpunct}
{\mcitedefaultendpunct}{\mcitedefaultseppunct}\relax
\EndOfBibitem
\bibitem[Paloli \emph{et~al.}(2013)Paloli, Mohanty, Crassous, Zaccarelli, and
  Schurtenberger]{schurtenberger2013}
D.~Paloli, P.~S. Mohanty, J.~J. Crassous, E.~Zaccarelli and P.~Schurtenberger,
  \emph{Soft Matter}, 2013, \textbf{9}, 3000--3004\relax
\mciteBstWouldAddEndPuncttrue
\mciteSetBstMidEndSepPunct{\mcitedefaultmidpunct}
{\mcitedefaultendpunct}{\mcitedefaultseppunct}\relax
\EndOfBibitem
\bibitem[Gasser \emph{et~al.}(2013)Gasser, Li{\'e}tor-Santos, Scotti, Bunk,
  Menzel, and Fern{\'a}ndez-Nieves]{nieves-pre2013}
U.~Gasser, J.-J. Li{\'e}tor-Santos, A.~Scotti, O.~Bunk, A.~Menzel and
  A.~Fern{\'a}ndez-Nieves, \emph{Phys. Rev. E}, 2013, \textbf{88},
  052308--1--8\relax
\mciteBstWouldAddEndPuncttrue
\mciteSetBstMidEndSepPunct{\mcitedefaultmidpunct}
{\mcitedefaultendpunct}{\mcitedefaultseppunct}\relax
\EndOfBibitem
\bibitem[Mohanty \emph{et~al.}(2014)Mohanty, Paloli, Crassous, Zaccarelli, and
  Schurtenberger]{schurtenberger2014}
P.~S. Mohanty, D.~Paloli, J.~J. Crassous, E.~Zaccarelli and P.~Schurtenberger,
  \emph{J. Chem. Phys.}, 2014, \textbf{140}, 094901--1--9\relax
\mciteBstWouldAddEndPuncttrue
\mciteSetBstMidEndSepPunct{\mcitedefaultmidpunct}
{\mcitedefaultendpunct}{\mcitedefaultseppunct}\relax
\EndOfBibitem
\bibitem[Braibanti \emph{et~al.}(2016)Braibanti, Haro-P\'erez, Quesada-P\'erez,
  Rojas-Ochoa, and Trappe]{braibanti-perez2016}
M.~Braibanti, C.~Haro-P\'erez, M.~Quesada-P\'erez, L.~F. Rojas-Ochoa and
  V.~Trappe, \emph{Phys. Rev. E}, 2016, \textbf{94}, 032601--1--8\relax
\mciteBstWouldAddEndPuncttrue
\mciteSetBstMidEndSepPunct{\mcitedefaultmidpunct}
{\mcitedefaultendpunct}{\mcitedefaultseppunct}\relax
\EndOfBibitem
\bibitem[Mason \emph{et~al.}(1995)Mason, Bibette, and Weitz]{weitz-prl1995}
T.~G. Mason, J.~Bibette and D.~A. Weitz, \emph{\PRL}, 1995, \textbf{75},
  2051--2054\relax
\mciteBstWouldAddEndPuncttrue
\mciteSetBstMidEndSepPunct{\mcitedefaultmidpunct}
{\mcitedefaultendpunct}{\mcitedefaultseppunct}\relax
\EndOfBibitem
\bibitem[Gr{\"o}hn and Antonietti(2000)]{groehn2000}
F.~Gr{\"o}hn and M.~Antonietti, \emph{Macromol.}, 2000, \textbf{33},
  5938--5949\relax
\mciteBstWouldAddEndPuncttrue
\mciteSetBstMidEndSepPunct{\mcitedefaultmidpunct}
{\mcitedefaultendpunct}{\mcitedefaultseppunct}\relax
\EndOfBibitem
\bibitem[Levin(2002)]{levin2002}
Y.~Levin, \emph{Phys. Rev. E}, 2002, \textbf{65}, 036143--1--6\relax
\mciteBstWouldAddEndPuncttrue
\mciteSetBstMidEndSepPunct{\mcitedefaultmidpunct}
{\mcitedefaultendpunct}{\mcitedefaultseppunct}\relax
\EndOfBibitem
\bibitem[Fern{\'a}ndez-Nieves and M{\'a}rquez(2005)]{nieves-jcp2005}
A.~Fern{\'a}ndez-Nieves and M.~M{\'a}rquez, \emph{J. Chem. Phys.}, 2005,
  \textbf{122}, 084702--1--6\relax
\mciteBstWouldAddEndPuncttrue
\mciteSetBstMidEndSepPunct{\mcitedefaultmidpunct}
{\mcitedefaultendpunct}{\mcitedefaultseppunct}\relax
\EndOfBibitem
\bibitem[Singh \emph{et~al.}(2012)Singh, Fedosov, Chatterji, Winkler, and
  Gompper]{winkler-gompper2012}
S.~P. Singh, D.~A. Fedosov, A.~Chatterji, R.~G. Winkler and G.~Gompper,
  \emph{J. Phys.: Condens. Matter}, 2012, \textbf{24}, 464103--1--11\relax
\mciteBstWouldAddEndPuncttrue
\mciteSetBstMidEndSepPunct{\mcitedefaultmidpunct}
{\mcitedefaultendpunct}{\mcitedefaultseppunct}\relax
\EndOfBibitem
\bibitem[Winkler \emph{et~al.}(2014)Winkler, Fedosov, and
  Gompper]{winkler-gompper2014}
R.~G. Winkler, D.~A. Fedosov and G.~Gompper, \emph{Curr. Opin. Colloid
  Interface Sci.}, 2014, \textbf{19}, 594--610\relax
\mciteBstWouldAddEndPuncttrue
\mciteSetBstMidEndSepPunct{\mcitedefaultmidpunct}
{\mcitedefaultendpunct}{\mcitedefaultseppunct}\relax
\EndOfBibitem
\bibitem[Ghavami and Winkler(2017)]{winkler2017}
A.~Ghavami and R.~G. Winkler, \emph{ACS Macro Lett.}, 2017, \textbf{6},
  721--725\relax
\mciteBstWouldAddEndPuncttrue
\mciteSetBstMidEndSepPunct{\mcitedefaultmidpunct}
{\mcitedefaultendpunct}{\mcitedefaultseppunct}\relax
\EndOfBibitem
\bibitem[Gnan \emph{et~al.}(2017)Gnan, Rovigatti, Bergman, and
  Zaccarelli]{zaccarelli2017}
N.~Gnan, L.~Rovigatti, M.~Bergman and E.~Zaccarelli, \emph{Macromol.}, 2017,
  \textbf{50}, 8777--8786\relax
\mciteBstWouldAddEndPuncttrue
\mciteSetBstMidEndSepPunct{\mcitedefaultmidpunct}
{\mcitedefaultendpunct}{\mcitedefaultseppunct}\relax
\EndOfBibitem
\bibitem[Li \emph{et~al.}(2014)Li, S{\'a}nchez-Di{\'a}z, Wu, Hamilton, Falus,
  Porcar, Liu, Do, Faraone, Smith, Egami, and Chen]{li-chen2014}
X.~Li, L.~E. S{\'a}nchez-Di{\'a}z, B.~Wu, W.~A. Hamilton, P.~Falus, L.~Porcar,
  Y.~Liu, C.~Do, A.~Faraone, G.~S. Smith, T.~Egami and W.-R. Chen, \emph{ACS
  Macro Lett.}, 2014, \textbf{3}, 1271--1275\relax
\mciteBstWouldAddEndPuncttrue
\mciteSetBstMidEndSepPunct{\mcitedefaultmidpunct}
{\mcitedefaultendpunct}{\mcitedefaultseppunct}\relax
\EndOfBibitem
\bibitem[Egorov \emph{et~al.}(2013)Egorov, Paturej, Likos, and
  Milchev]{egorov-likos2013}
S.~A. Egorov, J.~Paturej, C.~N. Likos and A.~Milchev, \emph{Macromol.}, 2013,
  \textbf{46}, 3648--3653\relax
\mciteBstWouldAddEndPuncttrue
\mciteSetBstMidEndSepPunct{\mcitedefaultmidpunct}
{\mcitedefaultendpunct}{\mcitedefaultseppunct}\relax
\EndOfBibitem
\bibitem[Colla \emph{et~al.}(2014)Colla, Likos, and Levin]{colla-likos2014}
T.~Colla, C.~N. Likos and Y.~Levin, \emph{J. Chem. Phys.}, 2014, \textbf{141},
  234902--1--11\relax
\mciteBstWouldAddEndPuncttrue
\mciteSetBstMidEndSepPunct{\mcitedefaultmidpunct}
{\mcitedefaultendpunct}{\mcitedefaultseppunct}\relax
\EndOfBibitem
\bibitem[Colla and Likos(2015)]{colla-likos2015}
T.~Colla and C.~N. Likos, \emph{Mol. Phys.}, 2015, \textbf{113},
  2496--2510\relax
\mciteBstWouldAddEndPuncttrue
\mciteSetBstMidEndSepPunct{\mcitedefaultmidpunct}
{\mcitedefaultendpunct}{\mcitedefaultseppunct}\relax
\EndOfBibitem
\bibitem[Colla \emph{et~al.}(2018)Colla, Mohanty, N{\"o}jd, Bialik, Riede,
  Schurtenberger, and Likos]{colla-likos2018}
T.~Colla, P.~S. Mohanty, S.~N{\"o}jd, E.~Bialik, A.~Riede, P.~Schurtenberger
  and C.~N. Likos, \emph{ACS Nano}, 2018,  DOI: 10.1021/acsnano.7b08843\relax
\mciteBstWouldAddEndPuncttrue
\mciteSetBstMidEndSepPunct{\mcitedefaultmidpunct}
{\mcitedefaultendpunct}{\mcitedefaultseppunct}\relax
\EndOfBibitem
\bibitem[Gupta \emph{et~al.}(2015)Gupta, Camargo, Stellbrink, Allgaier,
  Radulescu, Lindner, Zaccarelli, Likos, and
  Richter]{stellbrink-likos-nanoscale2015}
S.~Gupta, M.~Camargo, J.~Stellbrink, J.~Allgaier, A.~Radulescu, P.~Lindner,
  E.~Zaccarelli, C.~N. Likos and D.~Richter, \emph{Nanoscale}, 2015,
  \textbf{7}, 13924--13934\relax
\mciteBstWouldAddEndPuncttrue
\mciteSetBstMidEndSepPunct{\mcitedefaultmidpunct}
{\mcitedefaultendpunct}{\mcitedefaultseppunct}\relax
\EndOfBibitem
\bibitem[Gupta \emph{et~al.}(2015)Gupta, Stellbrink, Zaccarelli, Likos,
  Camargo, Holmqvist, Allgaier, Willner, and Richter]{stellbrink-likos-prl2015}
S.~Gupta, J.~Stellbrink, E.~Zaccarelli, C.~N. Likos, M.~Camargo, P.~Holmqvist,
  J.~Allgaier, L.~Willner and D.~Richter, \emph{Phys. Rev. Lett.}, 2015,
  \textbf{115}, 128302--1--5\relax
\mciteBstWouldAddEndPuncttrue
\mciteSetBstMidEndSepPunct{\mcitedefaultmidpunct}
{\mcitedefaultendpunct}{\mcitedefaultseppunct}\relax
\EndOfBibitem
\bibitem[Hedrick \emph{et~al.}(2015)Hedrick, Chung, and
  Denton]{hedrick-chung-denton2015}
M.~M. Hedrick, J.~K. Chung and A.~R. Denton, \emph{J. Chem. Phys.}, 2015,
  \textbf{142}, 034904--1--12\relax
\mciteBstWouldAddEndPuncttrue
\mciteSetBstMidEndSepPunct{\mcitedefaultmidpunct}
{\mcitedefaultendpunct}{\mcitedefaultseppunct}\relax
\EndOfBibitem
\bibitem[Urich and Denton(2016)]{urich-denton2016}
M.~Urich and A.~R. Denton, \emph{Soft Matter}, 2016, \textbf{12},
  9086--9094\relax
\mciteBstWouldAddEndPuncttrue
\mciteSetBstMidEndSepPunct{\mcitedefaultmidpunct}
{\mcitedefaultendpunct}{\mcitedefaultseppunct}\relax
\EndOfBibitem
\bibitem[Flory and Rehner(1943)]{flory-rehner1943-I}
P.~J. Flory and J.~Rehner, \emph{J. Chem. Phys.}, 1943, \textbf{11},
  512--520\relax
\mciteBstWouldAddEndPuncttrue
\mciteSetBstMidEndSepPunct{\mcitedefaultmidpunct}
{\mcitedefaultendpunct}{\mcitedefaultseppunct}\relax
\EndOfBibitem
\bibitem[Flory and Rehner(1943)]{flory-rehner1943-II}
P.~J. Flory and J.~Rehner, \emph{J. Chem. Phys.}, 1943, \textbf{11},
  521--526\relax
\mciteBstWouldAddEndPuncttrue
\mciteSetBstMidEndSepPunct{\mcitedefaultmidpunct}
{\mcitedefaultendpunct}{\mcitedefaultseppunct}\relax
\EndOfBibitem
\bibitem[Flory(1953)]{flory1953}
P.~J. Flory, \emph{Principles of Polymer Chemistry}, Cornell University Press,
  Ithaca, 1953\relax
\mciteBstWouldAddEndPuncttrue
\mciteSetBstMidEndSepPunct{\mcitedefaultmidpunct}
{\mcitedefaultendpunct}{\mcitedefaultseppunct}\relax
\EndOfBibitem
\bibitem[Landau and Lifshitz(1986)]{landau-lifshitz1986}
L.~D. Landau and E.~M. Lifshitz, \emph{Theory of Elasticity}, Elsevier,
  Amsterdam, 3rd edn., 1986\relax
\mciteBstWouldAddEndPuncttrue
\mciteSetBstMidEndSepPunct{\mcitedefaultmidpunct}
{\mcitedefaultendpunct}{\mcitedefaultseppunct}\relax
\EndOfBibitem
\bibitem[Denton(2003)]{denton2003}
A.~R. Denton, \emph{Phys. Rev. E}, 2003, \textbf{67}, 011804--1--10\relax
\mciteBstWouldAddEndPuncttrue
\mciteSetBstMidEndSepPunct{\mcitedefaultmidpunct}
{\mcitedefaultendpunct}{\mcitedefaultseppunct}\relax
\EndOfBibitem
\bibitem[Hu \emph{et~al.}(2011)Hu, Tong, and Lyon]{lyon-langmuir2011}
X.~Hu, Z.~Tong and L.~A. Lyon, \emph{Langmuir}, 2011, \textbf{27},
  4142--4148\relax
\mciteBstWouldAddEndPuncttrue
\mciteSetBstMidEndSepPunct{\mcitedefaultmidpunct}
{\mcitedefaultendpunct}{\mcitedefaultseppunct}\relax
\EndOfBibitem
\bibitem[Stieger \emph{et~al.}(2004)Stieger, Richtering, Pedersen, and
  Lindner]{stieger2004}
M.~Stieger, W.~Richtering, J.~S. Pedersen and P.~Lindner, \emph{J. Chem.
  Phys.}, 2004, \textbf{120}, 6197--6206\relax
\mciteBstWouldAddEndPuncttrue
\mciteSetBstMidEndSepPunct{\mcitedefaultmidpunct}
{\mcitedefaultendpunct}{\mcitedefaultseppunct}\relax
\EndOfBibitem
\bibitem[Moncho-Jord{\'a} \emph{et~al.}(2013)Moncho-Jord{\'a}, Anta, and
  Callejas-Fern{\'a}ndez]{moncho-jorda-anta2013}
A.~Moncho-Jord{\'a}, J.~A. Anta and J.~Callejas-Fern{\'a}ndez, \emph{J. Chem.
  Phys.}, 2013, \textbf{138}, 134902\relax
\mciteBstWouldAddEndPuncttrue
\mciteSetBstMidEndSepPunct{\mcitedefaultmidpunct}
{\mcitedefaultendpunct}{\mcitedefaultseppunct}\relax
\EndOfBibitem
\bibitem[Boon and Schurtenberger(2017)]{boon-schurtenberger2017}
N.~Boon and P.~Schurtenberger, \emph{Phys. Chem. Chem. Phys.}, 2017,
  \textbf{19}, 23740--23746\relax
\mciteBstWouldAddEndPuncttrue
\mciteSetBstMidEndSepPunct{\mcitedefaultmidpunct}
{\mcitedefaultendpunct}{\mcitedefaultseppunct}\relax
\EndOfBibitem
\bibitem[Quesada-P{\'e}rez and Mart{\'i}n-Molina(2013)]{quesada-perez2013}
M.~Quesada-P{\'e}rez and A.~Mart{\'i}n-Molina, \emph{Soft Matter}, 2013,
  \textbf{9}, 7086--7094\relax
\mciteBstWouldAddEndPuncttrue
\mciteSetBstMidEndSepPunct{\mcitedefaultmidpunct}
{\mcitedefaultendpunct}{\mcitedefaultseppunct}\relax
\EndOfBibitem
\bibitem[Adroher-Ben{\'i}tez \emph{et~al.}(2015)Adroher-Ben{\'i}tez, Ahualli,
  Mart{\'i}n-Molina, Quesada-P{\'e}rez, and
  Moncho-Jord{\'a}]{quesada-perez-moncho-jorda-anta2015}
I.~Adroher-Ben{\'i}tez, S.~Ahualli, A.~Mart{\'i}n-Molina, M.~Quesada-P{\'e}rez
  and A.~Moncho-Jord{\'a}, \emph{Macromol.}, 2015, \textbf{48},
  4645--4656\relax
\mciteBstWouldAddEndPuncttrue
\mciteSetBstMidEndSepPunct{\mcitedefaultmidpunct}
{\mcitedefaultendpunct}{\mcitedefaultseppunct}\relax
\EndOfBibitem
\bibitem[Rumyantsev \emph{et~al.}(2015)Rumyantsev, Rudov, and
  Potemkin]{potemkin2015}
A.~M. Rumyantsev, A.~A. Rudov and I.~I. Potemkin, \emph{J. Chem. Phys.}, 2015,
  \textbf{142}, 171105--1--5\relax
\mciteBstWouldAddEndPuncttrue
\mciteSetBstMidEndSepPunct{\mcitedefaultmidpunct}
{\mcitedefaultendpunct}{\mcitedefaultseppunct}\relax
\EndOfBibitem
\bibitem[Riest \emph{et~al.}(2015)Riest, Athanasopoulou, Egorov, Likos, and
  Ziherl]{riest2015}
J.~Riest, L.~Athanasopoulou, S.~A. Egorov, C.~N. Likos and P.~Ziherl,
  \emph{Sci. Rep.}, 2015, \textbf{5}, 15854--1--11\relax
\mciteBstWouldAddEndPuncttrue
\mciteSetBstMidEndSepPunct{\mcitedefaultmidpunct}
{\mcitedefaultendpunct}{\mcitedefaultseppunct}\relax
\EndOfBibitem
\bibitem[Cloitre and Bonnecaze(2010)]{cloitre-bonnecaze2010}
M.~Cloitre and R.~T. Bonnecaze, in \emph{High Solid Dispersions}, ed.
  M.~Cloitre, Springer, Heidelberg, 2010, pp. 117--161\relax
\mciteBstWouldAddEndPuncttrue
\mciteSetBstMidEndSepPunct{\mcitedefaultmidpunct}
{\mcitedefaultendpunct}{\mcitedefaultseppunct}\relax
\EndOfBibitem
\bibitem[Seth \emph{et~al.}(2011)Seth, Mohan, Locatelli-Champagne, Cloitre, and
  Bonnecaze]{cloitre-bonnacaze2011}
J.~R. Seth, L.~Mohan, C.~Locatelli-Champagne, M.~Cloitre and R.~T. Bonnecaze,
  \emph{Nature Mat.}, 2011, \textbf{10}, 838--843\relax
\mciteBstWouldAddEndPuncttrue
\mciteSetBstMidEndSepPunct{\mcitedefaultmidpunct}
{\mcitedefaultendpunct}{\mcitedefaultseppunct}\relax
\EndOfBibitem
\bibitem[Moncho-Jord{\'a} and Dzubiella(2016)]{moncho-jorda-dzubiella2016}
A.~Moncho-Jord{\'a} and J.~Dzubiella, \emph{Phys. Chem. Chem. Phys.}, 2016,
  \textbf{18}, 5372--5385\relax
\mciteBstWouldAddEndPuncttrue
\mciteSetBstMidEndSepPunct{\mcitedefaultmidpunct}
{\mcitedefaultendpunct}{\mcitedefaultseppunct}\relax
\EndOfBibitem
\bibitem[Lopez and Richtering(2017)]{richtering2017}
C.~G. Lopez and W.~Richtering, \emph{Soft Matter}, 2017, \textbf{13},
  8271--8280\relax
\mciteBstWouldAddEndPuncttrue
\mciteSetBstMidEndSepPunct{\mcitedefaultmidpunct}
{\mcitedefaultendpunct}{\mcitedefaultseppunct}\relax
\EndOfBibitem
\bibitem[Blundell and Terentjev(2009)]{blundell2009}
J.~R. Blundell and E.~M. Terentjev, \emph{Macromol.}, 2009, \textbf{42},
  5388--5394\relax
\mciteBstWouldAddEndPuncttrue
\mciteSetBstMidEndSepPunct{\mcitedefaultmidpunct}
{\mcitedefaultendpunct}{\mcitedefaultseppunct}\relax
\EndOfBibitem
\bibitem[Kim \emph{et~al.}(2017)Kim, Moncho-Jord{\'a}, Roa, Kandu{\v c}, and
  Dzubiella]{dzubiella2017}
W.~K. Kim, A.~Moncho-Jord{\'a}, R.~Roa, M.~Kandu{\v c} and J.~Dzubiella,
  \emph{Macromol.}, 2017, \textbf{50}, 6227--6237\relax
\mciteBstWouldAddEndPuncttrue
\mciteSetBstMidEndSepPunct{\mcitedefaultmidpunct}
{\mcitedefaultendpunct}{\mcitedefaultseppunct}\relax
\EndOfBibitem
\bibitem[de~Gennes(1979)]{deGennes1979}
P.-G. de~Gennes, \emph{Scaling Concepts in Polymer Physics}, Cornell, Ithaca,
  1979\relax
\mciteBstWouldAddEndPuncttrue
\mciteSetBstMidEndSepPunct{\mcitedefaultmidpunct}
{\mcitedefaultendpunct}{\mcitedefaultseppunct}\relax
\EndOfBibitem
\bibitem[Denton(2000)]{denton2000}
A.~R. Denton, \emph{Phys. Rev. E}, 2000, \textbf{62}, 3855--3864\relax
\mciteBstWouldAddEndPuncttrue
\mciteSetBstMidEndSepPunct{\mcitedefaultmidpunct}
{\mcitedefaultendpunct}{\mcitedefaultseppunct}\relax
\EndOfBibitem
\bibitem[Denton(2014)]{denton-cecam2014}
A.~R. Denton, in \emph{in Electrostatics of Soft and Disordered Matter}, ed.
  D.~S. Dean, J.~Dobnikar, A.~Naji and R.~Podgornik, Pan Stanford, Singapore,
  2014, pp. 201--215\relax
\mciteBstWouldAddEndPuncttrue
\mciteSetBstMidEndSepPunct{\mcitedefaultmidpunct}
{\mcitedefaultendpunct}{\mcitedefaultseppunct}\relax
\EndOfBibitem
\bibitem[Frenkel and Smit(2001)]{frenkel-smit2001}
D.~Frenkel and B.~Smit, \emph{Understanding Molecular Simulation}, Academic,
  London, 2nd edn., 2001\relax
\mciteBstWouldAddEndPuncttrue
\mciteSetBstMidEndSepPunct{\mcitedefaultmidpunct}
{\mcitedefaultendpunct}{\mcitedefaultseppunct}\relax
\EndOfBibitem
\bibitem[Binder and Heermann(2010)]{binder-heermann2010}
K.~Binder and D.~W. Heermann, \emph{{Monte} {Carlo} Simulation in Statistical
  Physics: {An} Introduction}, Springer, Berlin, 5th edn., 2010\relax
\mciteBstWouldAddEndPuncttrue
\mciteSetBstMidEndSepPunct{\mcitedefaultmidpunct}
{\mcitedefaultendpunct}{\mcitedefaultseppunct}\relax
\EndOfBibitem
\bibitem[van Roij and Hansen(1997)]{vanRoij1997}
R.~van Roij and J.~P. Hansen, \emph{Phys. Rev. Lett.}, 1997, \textbf{79},
  3082--3085\relax
\mciteBstWouldAddEndPuncttrue
\mciteSetBstMidEndSepPunct{\mcitedefaultmidpunct}
{\mcitedefaultendpunct}{\mcitedefaultseppunct}\relax
\EndOfBibitem
\bibitem[Denton(2006)]{denton2006}
A.~R. Denton, \emph{Phys. Rev. E}, 2006, \textbf{73}, 041407--1--14\relax
\mciteBstWouldAddEndPuncttrue
\mciteSetBstMidEndSepPunct{\mcitedefaultmidpunct}
{\mcitedefaultendpunct}{\mcitedefaultseppunct}\relax
\EndOfBibitem
\bibitem[Hansen and McDonald(2006)]{hansen-mcdonald2006}
J.-P. Hansen and I.~R. McDonald, \emph{Theory of Simple Liquids}, Elsevier,
  London, 3rd edn., 2006\relax
\mciteBstWouldAddEndPuncttrue
\mciteSetBstMidEndSepPunct{\mcitedefaultmidpunct}
{\mcitedefaultendpunct}{\mcitedefaultseppunct}\relax
\EndOfBibitem
\bibitem[Denton and Tang(2016)]{denton-tang2016}
A.~R. Denton and Q.~Tang, \emph{J. Chem. Phys.}, 2016, \textbf{145},
  164901--1--10\relax
\mciteBstWouldAddEndPuncttrue
\mciteSetBstMidEndSepPunct{\mcitedefaultmidpunct}
{\mcitedefaultendpunct}{\mcitedefaultseppunct}\relax
\EndOfBibitem
\bibitem[Quesada-P\'erez \emph{et~al.}(2018)Quesada-P\'erez, Maroto-Centeno,
  Mart\'{\i}n-Molina, and Moncho-Jord\'a]{quesada-perez-moncho-jorda2018}
M.~Quesada-P\'erez, J.~A. Maroto-Centeno, A.~Mart\'{\i}n-Molina and
  A.~Moncho-Jord\'a, \emph{Phys. Rev. E}, 2018, \textbf{97}, 042608--1--7\relax
\mciteBstWouldAddEndPuncttrue
\mciteSetBstMidEndSepPunct{\mcitedefaultmidpunct}
{\mcitedefaultendpunct}{\mcitedefaultseppunct}\relax
\EndOfBibitem
\bibitem[Hansen and Verlet(1969)]{hansen-verlet1969}
J.~P. Hansen and L.~Verlet, \emph{Phys. Rev.}, 1969, \textbf{184},
  151--161\relax
\mciteBstWouldAddEndPuncttrue
\mciteSetBstMidEndSepPunct{\mcitedefaultmidpunct}
{\mcitedefaultendpunct}{\mcitedefaultseppunct}\relax
\EndOfBibitem
\end{mcitethebibliography}

\providecommand*{\mcitethebibliography}{\thebibliography}
\csname @ifundefined\endcsname{endmcitethebibliography}
{\let\endmcitethebibliography\endthebibliography}{}

\end{document}